\documentclass[12pt]{article}
\usepackage{amsmath,amssymb}
\usepackage{bm}
\usepackage{color}
\usepackage[dvipsnames]{xcolor}
\usepackage{graphicx}
\usepackage{cite}
\usepackage{mathtools}
\usepackage{breqn}
\usepackage{braket, multirow, subfigure, multicol}
\usepackage{here}
\usepackage{slashed}
\usepackage{ulem}
\usepackage{comment}

\newcommand{\maru}[1]{\raise0.2ex\hbox{\textcircled{\scriptsize{#1}}}}

\numberwithin{equation}{section}

\begin{document}
\title{
\begin{flushright}
\ \\*[-80pt]
\begin{minipage}{0.23\linewidth}
\normalsize
EPHOU-24-004\\
KYUSHU-HET-289\\*[50pt]
\end{minipage}
\end{flushright}
{\Large \bf Spontaneous CP violation and \\ partially broken modular flavor symmetries 
\\*[20pt]}}

\author{
~Tetsutaro Higaki$^{a}$,
~Tatsuo Kobayashi$^{b}$, 
~Kaito Nasu$^{b}$ and
~Hajime Otsuka$^{c}$
\\*[20pt]
\centerline{
\begin{minipage}{\linewidth}
\begin{center}
{\it \normalsize
${}^{a}$Department of Physics, Keio University, Yokohama 223-8533, Japan \\
${}^{b}$Department of Physics, Hokkaido University, Sapporo 060-0810, Japan \\
${}^{c}$Department of Physics, Kyushu University, 744 Motooka, Nishi-ku, Fukuoka 819-0395, Japan
}
\\*[5pt]
\end{center}
\end{minipage}}
\\*[50pt]}

\date{
\centerline{\small \bf Abstract}
\begin{minipage}{0.9\linewidth}
\medskip
\medskip
\small
We study the realization of spontaneous CP violation through moduli stabilization. In modular flavor models, the source of CP violation is the vacuum expectation values of the complex structure moduli of toroidal compact space. We demonstrate that the combined effects of Type IIB flux compactifications with modular invariant couplings between the moduli and matter fields can induce spontaneous CP violation without or with supersymmetry breaking. Furthermore, some general properties of CP and modular invariant scalar potentials are presented. It is found that certain modifications or partial breakings of modular symmetry are useful in generating spontaneous CP violation. 
\end{minipage}
}
\begin{titlepage}
\maketitle
\thispagestyle{empty}
\end{titlepage}
\newpage


\section{Introduction}
\label{Intro}

The origin of CP violation is an important issue to study in particle physics.
The four-dimensional (4D) CP symmetry can be embedded into  
higher-dimensional proper Lorentz symmetry such as ten-dimensional (10D) 
proper Lorentz symmetry in superstring theory \cite{Green:2012pqa,Strominger:1985it,Dine:1992ya,Choi:1992xp}.
In such an underlying theory, the CP violation would occur spontaneously 
by moduli stabilization through compactification.
(See for early works on CP violation through moduli stabilization, e.g. 
Refs.~\cite{Acharya:1995ag,Dent:2001cc,Khalil:2001dr,Giedt:2002ns}.)

One of definite scenarios for moduli stabilization in the string theory
is flux compactifications \cite{Gukov:1999ya}.
Indeed, the possibility of spontaneous CP violation was studied 
by starting from CP symmetric three-form flux backgrounds in type IIB string theory 
on toroidal orientifolds such as $T^6/\mathbb{Z}_2$ and $T^6/(\mathbb{Z}_2 \times \mathbb{Z}'_2)$~\cite{Kobayashi:2020uaj}.
However, it was found that there remain flat directions, and CP-symmetric and 
CP-violating vacua are degenerate.
Such an analysis was extended to Calabi-Yau compactifications 
in Ref.~\cite{Ishiguro:2020nuf}.
The results are almost the same, although one exceptional example was shown, where the spontaneous CP-violating vacuum appears. 
Thus, the spontaneous CP violation hardly occurs at least on toroidal compactifications.

Torus and its orbifold compactifications as well as Calabi-Yau compactifications have the modular symmetries.
The modular symmetries are an important property in higher-dimensional theory on a compact space.
For example, the modular symmetry was studied in heterotic orbifold models \cite{Ferrara:1989qb,Lerche:1989cs,Lauer:1989ax,Lauer:1990tm}, and magnetized D-brane models on torus and its orbifold compactifications  \cite{Kobayashi:2018rad,Kobayashi:2018bff,Ohki:2020bpo,Kikuchi:2020frp,Kikuchi:2020nxn,
Kikuchi:2021ogn,Almumin:2021fbk,Kikuchi:2023awe}. 
(See also Refs.~\cite{Strominger:1990pd,Candelas:1990pi, Ishiguro:2020nuf, Ishiguro:2021ccl,Ishiguro:2024xph} for Calabi-Yau compactifications.) 
The three-form flux backgrounds break the modular symmetries, but sub-symmetries may remain \cite{Kobayashi:2020hoc}.

Recently, modular flavor symmetric models were studied intensively \cite{Feruglio:2017spp}.
(See for early works Refs.~~\cite{Kobayashi:2018vbk,Penedo:2018nmg,Criado:2018thu,Kobayashi:2018scp,Novichkov:2018ovf,Novichkov:2018nkm,deAnda:2018ecu,Okada:2018yrn,Kobayashi:2018wkl,Novichkov:2018yse} and for reviews Refs.~~\cite{Kobayashi:2023zzc,Ding:2023htn}.)
In these modular flavor symmetric models, quark and lepton masses, their mixing angles, and CP phases can be realized by choosing proper values of moduli,
although moduli values are treated as free parameters in many models.
For example, moduli values around fixed points can lead to hierarchical Yukawa 
couplings \cite{Feruglio:2021dte,Novichkov:2021evw,Petcov:2022fjf,Kikuchi:2023cap,Abe:2023ilq,Kikuchi:2023jap,Abe:2023qmr,Petcov:2023vws,Abe:2023dvr,deMedeirosVarzielas:2023crv,Kikuchi:2023dow,Kikuchi:2023fpl}.
At any rate, moduli are dynamical, and their vacuum expectation values (VEVs) are determined by moduli stabilization.
Indeed, moduli stabilization was studied in modular flavor symmetric models by assuming that the superpotential is a modular symmetric function \cite{Kobayashi:2019xvz,Kobayashi:2019uyt,Novichkov:2022wvg,Ishiguro:2022pde,Knapp-Perez:2023nty,King:2023snq,Kobayashi:2023spx}.
(See also for early works on modular symmetric superpotential Refs.~\cite{Font:1990nt,Ferrara:1990ei,Cvetic:1991qm}.
)
For example, it may be written by modular forms. 
In these approaches, moduli values leading to the successful flavor structure of quarks and leptons can be obtained, but the spontaneous CP violation hardly occurs.
One of the exceptional examples is the example studied in Ref.~\cite{Novichkov:2022wvg}.\footnote{
See also \cite{Knapp-Perez:2023nty} in which CP-breaking vacuum is obtained in the presence of matter field.}

Our purpose of this paper is to study the spontaneous CP violation 
through the moduli stabilization.
We study a CP invariant generic scalar potential, and implications of 
modular symmetry and its subgroup on the spontaneous CP violation.
We will show that the modular symmetry and its certain subgroup are 
important to preserve the CP symmetry. 
In order to discuss more concretely, 
we use models consisting of superpotential terms 
induced by the three-form flux and matter terms coupling to modular forms, which has been partly discussed in Refs.~\cite{Abe:2023ylh,Abe:2024tox}.
If such models satisfy certain conditions, 
the spontaneous CP violation can be realized.

This paper is organized as follows. 
In section \ref{sec:general}, we study generic properties of CP- and modular-invariant 
scalar potential. 
In section \ref{sec:Revew}, we study moduli stabilization 
by three-form flux backgrounds and matter fields coupled to
modular forms.
In sec. 3.1, we give a brief review of CP-symmetric superpotential, which is induced by three-form fluxes.
In sec. 3.2, we study the superpotential of matter fields coupled to
modular forms.
In section \ref{sec:model}, we combine the superpotential terms due to three-form fluxes and matter fields coupled to
modular forms in order to illustrate the spontaneous CP violation.
Section \ref{conclusion} is devoted to our conclusion.
In Appendix A, we show a proof on flux quanta leading to 
the spontaneous CP violation, which is discussed in Section 4.
In Appendix B, we give another model leading to 
the spontaneous CP violation, where the de Sitter vacuum is 
realized.
Modular forms of $A_4$ are reviewed in Appendix C.


\section{Properties of CP, modular invariant scalar potentials}
\label{sec:general}

Here, we show some general properties of a scalar potential $V$ which is symmetric under the CP and modular transformations. We consider $V$ as a function of single complex structure modulus $\tau$, $({\rm Im}(\tau)>0)$ of a toroidal compact space. The analysis performed in this section will be useful to understand why we typically obtain CP-conserving vacua under the control of modular symmetry.

\subsection{Extended modular symmetry}
The extended modular symmetry is a symmetry group generated by the CP and $\bar{\Gamma}=PSL(2,\mathbb{Z})$ modular transformations. The action of CP on $\tau$ is given by 
\begin{equation}
    \tau \xrightarrow{\rm CP} - \bar{\tau}.
\end{equation}
We note that CP transformation maps $\tau$ in the upper half plane $\{\tau \in \mathbb{C}|{\rm Im}(\tau)>0\}$ to the same plane. 
The action of $\bar{\Gamma}$ on $\tau$ is written as
\begin{equation}
    \tau \rightarrow \gamma \tau = \frac{a \tau + b}{c\tau + d},\quad \gamma \in \bar{\Gamma},
\end{equation}
where
\begin{equation}
    \gamma = 
    \begin{pmatrix}
    a & b \\ c & d    
    \end{pmatrix}, \quad a, b, c, d \in \mathbb{Z},\ ad -bc =1.
\end{equation}
The generators of $\bar{\Gamma}$ are given by 
\begin{equation}
    S = 
    \begin{pmatrix}
        0 & 1 \\ -1 & 0
    \end{pmatrix},\quad
    T =
    \begin{pmatrix}
        1 & 1 \\ 0 & 1
    \end{pmatrix}.
\end{equation}
Under $S$ and $T$, the modulus $\tau$ transforms as
\begin{equation}
    S:\ \tau \rightarrow - \frac{1}{\tau},\quad T:\ \tau \rightarrow \tau + 1.
\end{equation}
Using the above $S$- and $T$-transformations, any values of $\tau$ in the upper half plane can be mapped to the fundamental domain $\mathcal{D}$, which is roughly given by
\begin{equation}
\label{eq: fundamental_domain}
    \mathcal{D} = \left\{
    \tau \in \mathbb{C}:\  {\rm Im}(\tau) > 0,\ |{\rm Re}(\tau)| \leq \frac{1}{2},\ |\tau| \geq1    \right\}.
\end{equation}
The CP transformation can be combined with the $\bar{\Gamma}$ modular symmetry as an outer automorphism. The resulting group is known as the extended modular group:
\begin{equation}
\bar{\Gamma}^* = \left\langle \tau \xrightarrow{S} -1/ \tau,\ \tau \xrightarrow{T} \tau + 1,\ \tau \xrightarrow{{\rm CP}} - \bar{\tau}  \right\rangle,
\end{equation}
which is mathematically understood as a semi-direct product $\bar{\Gamma}^* \simeq \bar{\Gamma} \rtimes \mathbb{Z}_2^{\rm CP}$ \cite{Baur:2019kwi,Novichkov:2019sqv,Baur:2019iai,Baur:2020jwc,Nilles:2020gvu}.

\subsubsection{CP-conserving VEVs of $\tau$}
In modular flavor symmetric models, the vacuum expectation value of $\tau$ conserves CP when there exists a transformation $\gamma \in \bar{\Gamma}$ such that
\begin{equation}
    - \bar{\tau} = \gamma \tau.
\end{equation}
Since two vacua related by $\bar{\Gamma}$ are equivalent in a modular symmetric model, it is sufficient to consider VEVs inside the fundamental domain $\mathcal{D}$. By explicit analyses, it has been shown that the CP-conserving VEVs of $\tau$ are the imaginary axis ${\rm Re}(\tau)=0$ and the boundaries of the fundamental domain $\mathcal{D}$ \cite{Novichkov:2019sqv}.

\subsection{CP invariance}
We study the implications of CP symmetry in the scalar potential $V$.  By the CP invariance, we find
\begin{equation}
\label{eq: CP_Potential}
    V(-s,t) = V(s,t),
\end{equation}
where we have defined $s$ and $t$ as
\begin{equation}
    \tau = s + it.
\end{equation}
We immediately find 
\begin{equation}
      \frac{\partial V}{\partial s} (s=0,t) = 0,
\end{equation}
implying that local minimum or maximum of $V$ lie on the imaginary axis ${\rm Re}(\tau)=0$. 

\subsection{CP and $\bar{\Gamma}$ invariance}

By requiring $\bar{\Gamma}$ invariance to the CP invariant scalar potential $V$, directional derivatives of $V$ normal to the boundaries of the fundamental domain $\mathcal{D}$ will be shown to vanish there. 
Hence, on the boundaries of $\mathcal{D}$ or the imaginary axis ${\rm Re}(\tau)=0$, one finds local minimum or maximum of the scalar potential.
Since these regions correspond to CP-conserving points, one may expect that the potential minimum tends not to break the CP.
\footnote{There is however an example where CP-breaking vacua are realized by a CP symmetric modular invariant potential \cite{Novichkov:2022wvg}.} 
\subsubsection{CP and $T$ invariance}
From the $T$ and CP invariance, we obtain
\begin{equation}
    V(s,t) = V(s+1,t) = V(-s-1,t).
\end{equation}
Differentiating both sides with respect to $s$ yields
\begin{equation}
    \frac{\partial V}{\partial s}(s,t) = -   \frac{\partial V}{\partial s} (s',t),
\end{equation}
where $s':=-s-1$. By taking the limit $s \rightarrow -1/2$, we find
\begin{equation}
      \frac{\partial V}{\partial s} (s=-1/2,t) = 0.
\end{equation}
Noting that $V$ is periodic against the shift $s \rightarrow s+1$ and symmetric under $s \rightarrow -s$, we have
\begin{equation}
      \frac{\partial V}{\partial s} (s,t) = 0,\quad s \equiv 0 \pmod{1/2}.
\end{equation}

\subsubsection{CP and $S$ invariance}
From the $S$ and CP invariance, we obtain
\begin{equation}
    V(r,\theta) = V(1/r,-\theta + \pi) = V(1/r,\theta),
\end{equation}
where $\tau=r e^{i\theta}$.
Differentiating both sides with respect to $r$ yields
\begin{equation}
    \frac{\partial V}{\partial r}(r,\theta) = -  \frac{1}{r^2} \frac{\partial V}{\partial u} (u,\theta),
\end{equation}
where $u:=1/r$. By taking the limit $r \rightarrow 1$, we find\footnote{For example, in modular invariant inflation models, the modulus, which is an inflaton, slow-rolls along $r=1$ from $\tau=i(\omega)$ to $\tau=\omega(i)$ \cite{Kobayashi:2016mzg,Abe:2023ylh,Ding:2024neh,King:2024ssx}.}
\begin{equation}
      \frac{\partial V}{\partial r} (1,\theta) = 0.
\end{equation}

Note that finite fixed points, i.e. $\tau = \omega=e^{2 \pi i/3}~{\rm and}~i$, are the intersections of the CP invariant lines. Hence, the scalar potential $V$ is always stationary at the finite fixed points as pointed out in \cite{King:2023snq}, because two independent directional derivatives of $V$ vanish. 
This implies that we typically obtain 
vacuum stabilized at those fixed points if modular symmetry is present. When non-trivial contributions such as radiative corrections are considered, the VEV may deviate from the fixed points, however, it can hardly cause deviations from the CP-conserving lines \cite{Kobayashi:2023spx,Higaki:2024jdk} as long as modular symmetry is present.
In section \ref{sec:model} and Appendix \ref{appendix: Model-III}
, we will see that certain modifications or breakings of the modular symmetry can spontaneously induce CP-violating VEVs of moduli.

\subsection{CP and $\Gamma_0(N)$ invariance}
\label{sec:Gamma0(N)}

As an extension to the above analysis, we consider the $\Gamma_0(N)$ and the CP invariant scalar potential $V$. 
Here, $\Gamma_0(N)$ is defined as 
\begin{equation}
 \Gamma_0 (N)  = 
 \left\{ 
    \begin{pmatrix}
        a &  b \\
        c & d 
    \end{pmatrix} \in SL(2,\mathbb{Z}) \Bigg| \begin{pmatrix}
        a &  b  \\
        c & d 
    \end{pmatrix} \equiv
   \begin{pmatrix}
         * & \ * \\
        0 \ ({\rm mod}{N}) & * 
    \end{pmatrix} \right\},
\end{equation} 
where $*$ denotes unspecified integers.
For the moment, we focus on two 
generators for $\Gamma_0(N)$:
\begin{align}
T=
\left(
\begin{array}{cc}
    1 & 1 \\
    0 & 1
\end{array}
\right)
\,
,
\qquad
U =
\left(
\begin{array}{cc}
    1 & 0 \\
    N & 1
\end{array}
\right)\,.
\end{align}

First, we focus on the invariance of the scalar potential under CP and $T$:
\begin{align}
V(s,t) = V(s+1, t) = V(-s-1,t).    
\end{align}
In a similar analysis with the $SL(2,\mathbb{Z})$ case, the stationary point of the scalar potential is 
\begin{equation}
      \frac{\partial V}{\partial s} (s,t) = 0,\quad s \equiv 0 \pmod{1/2}.
\end{equation}
Next, we consider the scalar potential which is invariant under the $U$ transformation: 
\begin{align}
    \tau \xrightarrow{U} \tau^\prime=\frac{\tau}{N\tau +1}.
\end{align}
Acting the $U$ and CP transformations on $\tau= -\frac{1}{N} + r e^{i\theta}$, it turns out that
\begin{align}
    \tau \xrightarrow{U} \frac{1}{N} - \frac{1}{N^2}\frac{e^{-i\theta}}{r}
    \xrightarrow{CP} - \left( \frac{1}{N} - \frac{1}{N^2}\frac{e^{-i\theta}}{r}\right)^\ast
    = -\frac{1}{N} + \frac{1}{N^2} \frac{e^{i\theta}}{r}\,.
\end{align}
Then, we arrive at
\begin{equation}
    V(r,\theta) = V\left(\frac{1}{N^2r},\theta\right).
\end{equation}
By differentiating both sides with respect to $r$ and taking the limit $r\rightarrow 1/N$, 
we find
\begin{equation}
      \frac{\partial V}{\partial r} \left(\frac{1}{N},\theta \right) = 0.
\end{equation}

Hence, the scalar potential $V$ is always stationary at the intersection point of CP-conserving arc, i.e., $|\tau + \frac{1}{N}|=\frac{1}{N}$ and ${\rm Re}(\tau)\equiv 0\pmod{1/2}$, which includes the fixed point of $\Gamma_0(N)$, as will be shown later.\footnote{The arc $|\tau + \frac{1}{N}|=\frac{1}{N}$ is CP-conserving, because $-\bar{\tau}=\gamma \tau$ is satisfied by choosing $\gamma = S T^{-N} S$.} 
In the following, we discuss a specific $\Gamma_0(N)$ and check whether the fixed point of $\Gamma_0(N)$ is stationary.\footnote{It was recently pointed out that such a modular symmetry can appear in the low-energy effective theory of string theory \cite{Ishiguro:2023wwf,Ishiguro:2024xph}.}

\paragraph{CP and $\Gamma_0(2)$ invariance}\,\\

We choose generators for $\Gamma_0(2)$:
\begin{align}
T=
\left(
\begin{array}{cc}
    1 & 1 \\
    0 & 1
\end{array}
\right)
\,
,
\qquad
S^\prime =
\left(
\begin{array}{cc}
    -1 & 1 \\
    -2 & 1
\end{array}
\right)\,,
\end{align}
where the order of $S^\prime = S^{-1}T^{-2}S^{-1}T^{-1}$ is 4, i.e., 
$(S^\prime)^2= -\mathbf{1}$.
The fixed point under $S^\prime$ is located at $\tau_\ast = \frac{1+i}{2}$. 
Note that the generators $U$ and $S^\prime$ are related as $U=- S^\prime T$. 
Indeed, one can check that the scalar potential is stationary at the fixed point under $S^\prime$.

\paragraph{CP and $\Gamma_0(3)$ invariance}\,\\

We choose generators for $\Gamma_0(3)$:
\begin{align}
T=
\left(
\begin{array}{cc}
    1 & 1 \\
    0 & 1
\end{array}
\right)
\,
,
\qquad
S^\prime =
\left(
\begin{array}{cc}
    -1 & 1 \\
    -3 & 2
\end{array}
\right)\,,
\end{align}
where the order of $S^\prime= S^{-1}T^{-3}S^{-1}T^{-1}$ is 6, i.e., 
$(S^\prime)^3= -\mathbf{1}$. 
The fixed point under $S^\prime$ is located at $\tau_\ast = \frac{1}{2} + i \frac{\sqrt{3}}{6}$. 
Note that the generators $U$ and $S^\prime$ are related as $U=- S^\prime T$. 
At the fixed point under $S^\prime$, that is, $\tau=\tau_\ast$, 
we find that the scalar potential is stationary. 

\paragraph{CP and $\Gamma_0(5)$ invariance}\,\\

We choose generators for $\Gamma_0(5)$:
\begin{align}
T=
\left(
\begin{array}{cc}
    1 & 1 \\
    0 & 1
\end{array}
\right)
\,
,
\qquad
S_1 =
\left(
\begin{array}{cc}
    2 & -1 \\
    5 & -2
\end{array}
\right)\,,
\qquad
S_2 =
\left(
\begin{array}{cc}
    3 & -2 \\
    5 & -3
\end{array}
\right)\,,
\end{align}
where the order of $S_1$ and $S_2$ is 4, i.e., 
$(S_1)^2= (S_2)^2= -\mathbf{1}$. 
The fixed point under $S_1$ and $S_2$ are located at
$\tau_*^{(1)} = \frac{2+i}{5}$ and $\tau_*^{(2)} = \frac{3+i}{5}$, respectively. 
Both $\tau_*^{(1)}$ and $\tau_*^{(2)}$ are CP-conserving points.
Note that the generators $U$, $S_1$ and $S_2$ are related as $U=- S_1^{-1}S_2^{-1} T$. 
At the fixed points of
$\tau=\tau_*^{(1)}$ and $\tau_*^{(2)}$, 
we find that the scalar potential is stationary. 

Similarly, we can study the fixed points and scalar potential behavior for other $\Gamma_0(N)$.\footnote{
$\Gamma_0(4)$ has three cusps at $\tau= \infty, 0$ and $-1/2$. These values are not realistic for compactifications since ${\rm Im}\tau =0$. If the imaginary part of $\tau$ is vanishing, toroidal space such as 2D torus $T^2$ is reduced to an one-dimensional sphere $S^1$. Since the dimension of the compact space such as $T^6/\mathbb{Z}_2 = (T^2 \times T^2 \times T^2)/\mathbb{Z}_2$ should be conserved at 6, those modulus VEVs are undesirable.}

\section{Theoretical setup}
\label{sec:Revew}
Here, we construct CP invariant moduli potential. We will first review flux compactification. After that, we introduce a matter field coupled symmetrically to a modulus under the extended modular group. 

\subsection{CP invariant flux compactifications}
Here, we give a review of CP symmetric superpotential, 
which is induced by three-form fluxes \cite{Kobayashi:2020uaj}.
That leads to flat directions, where CP-conserving and 
CP-violating vacua are degenerate.

We study the low-energy effective field theory of Type IIB string on the factorizable $T^6 / \mathbb{Z}_2 = (T_1^2 \times T_2^2 \times T_2^3)/\mathbb{Z}_2$ orientifold. Here and hereafter, we adopt the reduced Planck mass unit $M_{\rm Pl}=1$. We consider the 4D $\mathcal{N}=1$ supergravity (SUGRA) action. The K\"{a}hler potential $K_{\rm moduli}$ is given by 
\begin{equation}
\label{eq: Kahler_flux}
    K_{\rm moduli} = - \ln{[-i(S-\bar{S})]} - \ln{[i(\tau_1-\bar{\tau}_1)(\tau_2-\bar{\tau}_2)(\tau_3-\bar{\tau}_3)]} -2 \ln{\mathcal{V}},
\end{equation}
where $S$, $\tau_i$ and $\mathcal{V}$ denote the axio-dilaton, complex structure moduli and the K\"{a}hler (volume) moduli, respectively.
In the presence of background three-form flux $G_3$, we have the Gukov-Vafa-Witten superpotential \cite{Gukov:1999ya}:
\begin{equation}
\label{eq: GVW}
    W_{\rm flux} = \frac{1}{l_s^2} \int  G_3 \wedge \Omega ,
\end{equation}
where $l_s=2\pi \sqrt{\alpha'}$ denotes the string length. The holomorphic three-form $\Omega$ is constructed by 
\begin{equation}
\label{eq: Omega}
    \Omega = dz_1 \wedge dz_2 \wedge dz_3,
\end{equation}
where $dz_i = dx^i + \tau_i dy^i, (i=1,2,3)$. The three-form flux $G_3$ is written by the linear combinations of Ramond-Ramond (R-R) three-form $F_3$ and the Neveu-Schwarz (NS-NS) three-form $H_3$ as
\begin{equation}
    G_3 = F_3 - S H_3.
\end{equation}
We expand $F_3$ and $H_3$ by the basis of $H^3(T^6, \mathbb{Z})$ satisfying $\int_{T^6} \alpha_I \wedge \beta^J = \delta_{I}^J$:
\begin{align}
\begin{aligned}
\frac{1}{l_s^2} F_3 &= a^0 \alpha_0 + a^i \alpha_i + b_i \beta^i + b_0 \beta^0, \\  
\frac{1}{l_s^2} H_3 &= c^0 \alpha_0 + c^i \alpha_i + d_i \beta^i + d_0 \beta^0,
\end{aligned}
\end{align}
where $a^0, a^i, b_0, b_i, c^0, c^i, d_0, d_i$ are all quantized to integers:
\begin{align}
    \frac{1}{l_s^2} \int_{\Sigma_3} F_3 \in \mathbb{Z},\quad \frac{1}{l_s^2} \int_{\Sigma_3} H_3 \in \mathbb{Z},
\end{align}
for any 3-cycle $\Sigma_3$. 
The flux quanta can take odd integers if we assume exotic $O3'$-planes in the system
\cite{Hanany:2000fq, Witten:1997bs}.
We can construct the basis explicitly as
\begin{align}
\begin{aligned}
    \alpha_0 &= dx^1 \wedge dx^2 \wedge dx^3,\quad &\alpha_1 = dy^1 \wedge dx^2 \wedge dx^3, \\
    \alpha_2 &= dy^2 \wedge dx^3 \wedge dx^1,\quad &\alpha_3 = dy^3 \wedge dx^1 \wedge dx^2, \\
    \beta_0 &= dy^1 \wedge dy^2 \wedge dy^3,\quad &\beta_1 = -dx^1 \wedge dy^2 \wedge dy^3, \\
    \beta_2 &= -dx^2 \wedge dy^3 \wedge dy^1,\quad &\beta_3 = -dx^3 \wedge dy^1 \wedge dy^2.
\end{aligned}
\end{align}
We note that the size of flux quanta cannot be arbitrarily large owing to the tadpole cancellation condition:
\begin{align}
\begin{aligned}
\label{eq: tadpole}
n_{\rm flux} &= \frac{1}{l_s^2} \int H_3 \wedge F_3 \\
&= c^0 b_0 - d_0 a^0 + \sum_i (c^i b_i - d_i a^i) \\
&= 32 - 2n_{\rm D3} - n_{\rm O3'} \leq 32,
\end{aligned}
\end{align}
where $n_{\rm D3}$ and $n_{\rm O3'}$ denote the number of $D3$-branes and exotic $O3'$-planes, respectively.
Furthermore, the imaginary self-duality condition on the flux $\star_6 G_3 = iG_3$ 
requires $n_{\rm flux} \geq 0$,
where $\star_6$ is the Hodge dual on the six dimensional extra dimension.

\subsubsection{CP transformation}
The 4D CP is embedded into the 10D proper Lorentz transformation. The complex coordinates of the 6D extra dimensions are transformed as $z_i \rightarrow -\bar{z}_i,\ (i=1,2,3)$ and the 4D parity is simultaneously performed under the CP transformation. As a result, the complex structure moduli and the axio-dilaton are transformed as
\begin{equation}
    \tau_i \xrightarrow{{\rm CP}} - \bar{\tau}_i,\quad S \xrightarrow{{\rm CP}} - \bar{S}.
\end{equation}
The holomorphic three-form behaves as
\begin{equation}
\label{eq: CP_Omega}
\Omega \xrightarrow{{\rm CP}} - \bar{\Omega}.
\end{equation}
Since the inversion $z_i \rightarrow -\bar{z}_i$ corresponds to $x_i \rightarrow -x_i$ and $y_i \rightarrow y_i$ in real coordinates, the basis of $H^3(T^6,\mathbb{Z})$ are transformed as
\begin{align}
\label{eq: basis_CP}
    \alpha_0 \xrightarrow{\rm CP} - \alpha_0,\quad \beta^0 \xrightarrow{\rm CP} \beta^0, \quad \alpha_i  \xrightarrow{\rm CP}  \alpha_i, \quad \beta^i  \xrightarrow{\rm CP} - \beta^i.
\end{align}

\subsubsection{CP invariant potential}
To construct a CP invariant scalar potential in the SUGRA, the flux-induced superpotential (\ref{eq: GVW}) needs to transform as 
\begin{equation}
\label{eq: W->barW}
W_{\rm flux} \xrightarrow{\rm CP}  e^{i \gamma} \bar{W}_{\rm flux},
\end{equation}
where $\gamma \in \mathbb{R}$. In fact, $\gamma$ is restricted to $\gamma = 0, \pi \pmod{2\pi}$ as follows. 
We first note that the orientation of the orientifold cover is reversed by the CP transformation. We denote it by $\mathcal{M}(=T^6) \xrightarrow{\rm CP} \mathcal{M}_{(\rm CP)}$. This corresponds to the sign difference between $\int_{\mathcal{M}} \alpha^I \wedge \beta_J = \delta^I_J$ and $\int_{\mathcal{M}_{(CP)}} \alpha^I \wedge \beta_J =- \delta^I_J$. Taking the orientation reversing into consideration along with eq.~(\ref{eq: CP_Omega}), we require
\begin{equation}
\label{eq: G3}
    G_3 \xrightarrow{{\rm CP}}   e^{i \gamma} \bar{G}_3,
\end{equation}
to realize eq.~(\ref{eq: W->barW}). 
We can translate eq.~(\ref{eq: G3}) to
\begin{align}
\begin{aligned}
\label{eq: CP_F_H}
    F_3  &\xrightarrow{\rm CP}  e^{i \gamma} F_3, \\
    H_3  &\xrightarrow{\rm CP} - e^{i \gamma} H_3. 
\end{aligned}
\end{align}
Since $H_3$ and $F_3$ are real three-forms, $e^{i\gamma} \in \mathbb{R}$ is required. 

Let us first choose $\gamma=\pi \pmod{2\pi}$.
To satisfy eq.~(\ref{eq: CP_F_H}), the $F_3$ and $H_3$ are restricted to
\begin{align}
\label{eq: gamma=0_F3}
   \frac{1}{l_s^2} F_3 &= a^0 \alpha_0  + b_i \beta^i,\\
\label{eq: gamma=0_H3}
   \frac{1}{l_s^2} H_3 &= c^i \alpha_i  + d_0 \beta^0, 
\end{align}
by taking into account eq.~(\ref{eq: basis_CP}).
As a result, one obtains the following superpotential:
\begin{equation}
\label{eq: W_flux_CP_invariant}
    W_{\rm flux} = a^0 \tau_1 \tau_2  \tau_3 + c^1 S \tau_2 \tau_3 + c^2 S \tau_1 \tau_3 + c^3 S \tau_1 \tau_2 - \sum_{i=1}^3 b_i \tau_i + d_0 S.
\end{equation}
One can directly see that the superpotential (\ref{eq: W_flux_CP_invariant}) behaves as $W_{\rm flux} \xrightarrow{\rm CP} - \bar{W}_{\rm flux}$ because $W_{\rm flux}$ is an odd polynomial of the moduli fields. In this paper, we understand that all the flux quanta $a^0, b_i, c^i, d_0$ in eq.~(\ref{eq: W_flux_CP_invariant}) should be invariant under the CP otherwise they explicitly break the CP symmetry as spurions. As a consistency check, let us focus on one of the flux quanta $a^0 = \int_{\mathcal{M}} F_3 \wedge \beta_0$. If we perform the CP transformation, we obtain $\int_{\mathcal{M}_{(\rm CP)}} (-F_3) \wedge \beta_0 = -a^0 \int_{\mathcal{M}_{(\rm CP)}} \alpha^0 \wedge \beta_0 = a^0$ as it should be. 

On the other hand, if we choose $\gamma =0 \pmod{2\pi}$, the $F_3$ and $H_3$ are restricted to
\begin{align}
   \frac{1}{l_s^2} F_3 &= a^i \alpha_i  + b_0 \beta^0,\\
   \frac{1}{l_s^2} H_3 &= c^0 \alpha_0  + d_i \beta^i.  
\end{align}
As a result, one obtains the even polynomial of moduli fields,
\begin{align}
\label{eq: W_even}
   W_{\rm flux} = -c^0 S\tau_1\tau_2\tau_3 - a^1 \tau_2\tau_3
    - a^2\tau_1\tau_3 - a^3\tau_1\tau_2 + \sum_{i=1}^3 d_i S \tau_i -b_0 ,
\end{align}    
which behaves as $W_{\rm flux} \xrightarrow{\rm CP} \bar{W}_{\rm flux}$. The flux quanta in eq.~(\ref{eq: W_even}) are also CP invariant.

\subsubsection{Flat directions}
Here, we focus on the superpotential of the form eq.~(\ref{eq: W_flux_CP_invariant}).
Following \cite{Hebecker:2017lxm, Kobayashi:2020hoc}, we restrict the flux quanta by 
\begin{equation}
\label{eq: Restricted_Flux_Quanta}
    c^1 = c^2=0,\quad b_1 = b_2 =0,\quad c^3 = - f a^0,\quad d_0=fb_3.
\end{equation}
By definition, $f$ is a rational number. Then, the CP-conserving superpotential (\ref{eq: W_flux_CP_invariant}) is reduced to 
\begin{equation}
\label{eq: simple_W_flux}
    W_{\rm flux}=(\tau_3-fS)\left[a^0 \tau_1 \tau_2 - b_3 \right].
\end{equation}
The potential minimum corresponds to 
\begin{equation}
    \partial_{\tau_i} W_{\rm flux} =0,\quad \partial_{S} W_{\rm flux} =0, \quad W_{\rm flux} = 0,
\end{equation}
which is given by 
\begin{equation}
\label{eq: flux_model2}
    \tau_1 \tau_2 = \frac{b_3}{a^0},\quad  \tau_3 = fS,
\end{equation}
where supersymmetry (SUSY) remains unbroken.
This shows that there exist flat directions in the moduli space where CP-conserving and violating vacua are degenerate.\footnote{
There can exist similar flat directions even when the superpotential in eq.~(\ref{eq: W_even}) is chosen.
For example, see Model 1 of Ref.~\cite{Kobayashi:2020hoc}.}
Note that the flux-induced D3-brane charge is given by 
\begin{equation}
    n_{\rm flux} = -2 f a^0 b_3.
\end{equation}
Hence, the tadpole cancellation condition eq.~(\ref{eq: tadpole}) and the imaginary self-duality condition require
\begin{equation}
\label{eq: tadpole2}
    0 \leq -fa^0 b_3 \leq 16.
\end{equation}
Furthermore, the flux quanta are constrained by
\begin{equation}
    {\rm sgn}(f) = +1, 
\end{equation}
from eq.~(\ref{eq: flux_model2}) because we restrict to ${\rm Im}(\tau_i)>0,\ {\rm Im}(S)>0$ in our convention. Then, we immediately notice ${\rm sgn}(a^0 b_3)=-1$.

\subsubsection{Moduli masses}
Here, we analyze the moduli masses by computing the Hessian of the SUGRA scalar potential:
\begin{equation}
    V = e^{K_{\rm moduli}} K^{A \bar{B}} D_A W_{\rm flux} \overline{D_B W_{\rm flux}},
\end{equation}
where $A, B \in \{ \tau_i, S \}$ and $K^{A\bar{B}}$ denotes the inverse of the K\"{a}hler metric $K_{A\bar{B}}= \partial_A \partial_{\bar{B}} K_{\rm moduli}$. It is assumed that the K\"ahler moduli have no-scale type K\"ahler potential $-2\ln {\cal V}$ which cancels out the supergravity effect $-3|W_{\rm flux}|^2$.
The covariant derivative of the superpotential is defined as 
\begin{equation}
    D_A W_{\rm flux} = \partial_A W_{\rm flux} + (\partial_A K_{\rm moduli}) W_{\rm flux}.
\end{equation}
At the SUSY vacuum, i.e. $\partial_A W_{\rm flux}= W_{\rm flux}=0$, the Hessian is given by
\begin{align}
\label{eq: Hessian_Flux}
    H = 
    \begin{pmatrix}
    \partial_A \bar{\partial}_{\bar{B}} V & \partial_A \partial_{B} V \\
    \bar{\partial}_{\bar{A}} \bar{\partial}_{\bar{B}} V &   \bar{\partial}_{\bar{A}} \partial_B V 
    \end{pmatrix} = 
     \langle e^{ K_{\rm moduli} }\rangle
    \begin{pmatrix}
        m_{AC} K^{C\bar{D}} \bar{m}_{\bar{D}\bar{B}} & 0 \\ 0 & \bar{m}_{\bar{A}\bar{C}} K^{\bar{C} D} m_{DB}
    \end{pmatrix},
\end{align}
where $m_{AB}$ denotes the supersymmetric mass:
\begin{equation}
\label{eq: SUSY_mass_def}
    m_{AB} = \partial_A \partial_B W_{\rm flux}.
\end{equation}
Non-vanishing components are given by 
\begin{align}
\begin{aligned}
    m_{\tau_1 \tau_3} &= m_{\tau_3 \tau_1} = a^0 \langle \tau_2 \rangle, \\
    m_{\tau_1 S} &= m_{S \tau_1} = - f a^0 \langle \tau_2 \rangle, \\
    m_{\tau_2 \tau_3} &= m_{\tau_3 \tau_2} = a^0 \langle \tau_1 \rangle, \\
    m_{\tau_2 S} &= m_{S \tau_2} = - f a^0 \langle \tau_1 \rangle.
\end{aligned}
\end{align}
Hence, the top-left $4 \times 4$ block of the Hessian matrix in eq.~(\ref{eq: Hessian_Flux})
is given by
\begin{align}
\begin{aligned}
\label{eq: Hessian_Flux_top-left}
\begin{pmatrix}
    H_{\tau_1 \bar{\tau}_1} & H_{\tau_1 \bar{\tau}_2} & H_{\tau_1 \bar{\tau}_3} & H_{\tau_1 \bar{S}} \\
    H_{\tau_2 \bar{\tau}_1} & H_{\tau_2 \bar{\tau}_2} & H_{\tau_2 \bar{\tau}_3} & H_{\tau_2 \bar{S}} \\
    H_{\tau_3 \bar{\tau}_1} & H_{\tau_3 \bar{\tau}_2} & H_{\tau_3 \bar{\tau}_3} & H_{\tau_3 \bar{S}} \\ 
    H_{S \bar{\tau}_1} & H_{S \bar{\tau}_2} & H_{S \bar{\tau}_3} & H_{S \bar{S}}
\end{pmatrix}
 = \langle e^{ K_{\rm moduli} }\rangle
\begin{pmatrix}
|\langle \tau_2 \rangle|^2 \zeta & \langle \bar{\tau}_1 \rangle \langle \tau_2 \rangle \zeta & 0 & 0 \\
\langle \tau_1 \rangle \langle \bar{\tau}_2 \rangle \zeta & |\langle \tau_1 \rangle|^2 \zeta  & 0 & 0 \\
0 & 0 & (a^0)^2 \eta & - f (a^0)^2 \eta \\
0 & 0 & - f (a^0)^2 \eta & (fa^0)^2 \eta
\end{pmatrix},
\end{aligned}
\end{align}
where 
\begin{align}
\begin{aligned}
    \zeta &= (a^0)^2 [2{\rm Im}\langle \tau_3 \rangle]^2 + (f a^0)^2 [2{\rm Im}\langle S \rangle]^2, \\
    \eta &= [2{\rm Im}\langle \tau_1 \rangle]^2 |\langle \tau_2 \rangle|^2 + [2{\rm Im}\langle \tau_2 \rangle]^2 |\langle \tau_1 \rangle|^2.
\end{aligned}
\end{align}
By diagonalizing the Hessian matrix in eq.~(\ref{eq: Hessian_Flux_top-left}), we find
\begin{equation}
\label{eq: Mass_scale_Flux}
\langle e^{ K_{\rm moduli} }\rangle
    \begin{pmatrix}   
(|\langle \tau_1 \rangle|^2 +|\langle \tau_2 \rangle|^2) \zeta & 0 & 0 & 0 \\
0 & 0 & 0 & 0 \\
0 & 0 & [(a^0)^2 + (fa^0)^2] \eta  & 0 \\
0 & 0 & 0 & 0 
    \end{pmatrix}.
\end{equation}
This shows that only half of the moduli can acquire masses.
The appearance of the zero-eigenvalues corresponds to the existence of the flat directions.

\subsection{Matter contribution}
We need additional sources to resolve the degeneracy between CP-conserving and breaking vacua produced by the flux-induced potential. As one of the candidates, we consider matter fields coupled to the moduli. We require the matter contributions to conserve the extended modular symmetry. In this subsection, we will focus on the matter contribution in the presence of only one modulus $\tau$ for simplicity. After briefly reviewing modular forms, we will show the 4D $\mathcal{N}=1$ SUGRA Lagrangian corresponding to the matter contributions. Then, we give illustrating examples using $A_4$ modular forms. We will see that CP is hardly violated solely by the matter contribution owing to the modular symmetry as studied in Section~\ref{sec:general}.
Discussions will be developed for the multiple moduli case in the next section where we consider the combined effects with flux-induced potential in the hope of spontaneous CP violation.

\subsubsection{Modular forms}
We briefly review modular forms which play a crucial role when
writing down modular symmetric modulus-matter couplings.
We begin with the following infinite subgroup of $\bar{\Gamma}$:
\begin{equation}
    \bar{\Gamma}(N) = \left\{ 
    \begin{pmatrix}
    a & b \\ c  & d 
    \end{pmatrix}
     \in \bar{\Gamma} \  \Bigg|  
    \begin{pmatrix}
    a & b \\ c  & d 
    \end{pmatrix}
      \equiv
    \begin{pmatrix}
    1 & 0 \\ 0  & 1 
    \end{pmatrix}
    \pmod{N} \right\} ,
\end{equation}
where $N$ is a positive integer. 
The finite modular subgroups $\Gamma_N$ are defined by the quotient $\Gamma_N \equiv \bar{\Gamma}/\bar{\Gamma}(N)$. 
It is known that $\Gamma_N$ is isomorphic to the non-Abelian discrete symmetry groups $S_3, A_4, S_4, A_5$ for $N=2,3,4,5$ respectively \cite{deAdelhartToorop:2011re}.

The modular forms $Y_i(\tau)$ of weight $k_Y$ and level $N$ are holomorphic functions, which transform under $\bar{\Gamma}$ as
\begin{equation}
    Y_i(\tau) \rightarrow (c \tau + d)^{k_Y} \rho(\gamma)_{ij} Y_j(\tau), \quad  \begin{pmatrix}
    a & b \\ c  & d 
    \end{pmatrix} \in \bar{\Gamma},
\end{equation}
where $\rho(\gamma)$ denotes a unitary representation matrix of $\gamma \in \Gamma_N$.

\subsubsection{4D $\mathcal{N}=1$ SUGRA Lagrangian}
The K\"{a}hler potential is given by 
\begin{equation}
    K = - \ln[-i(\tau - \bar{\tau})] + Z |X|^2- 2\ln{\mathcal{V}}, \quad Z = (-i\tau - i \bar{\tau})^k.
\end{equation}
We consider the simplest case where $Y(\tau)$ belongs to a trivial singlet representation $\bf{1}$ of $\Gamma_N$, i.e.
\begin{equation}
    Y(\tau) \rightarrow (c\tau + d)^{k_Y} Y(\tau),\quad \gamma \in \bar{\Gamma}.
\end{equation}
We denote the matter field by $X$, and it carries a modular weight $k \in \mathbb{Z}$,  
\begin{equation}
    X \rightarrow (c \tau + d)^k X,\quad \gamma \in \bar{\Gamma}.
\end{equation}
We have assumed that $X$ also belongs to the trivial singlet $\bf{1}$. We study the following superpotential: 
\begin{equation}
    W_{\rm matter} = \Lambda^2 Y(\tau) X,
\end{equation}
where $\Lambda$ is a mass parameter characterizing the coupling between the modulus $\tau$ and the matter $X$ \cite{Abe:2023ylh,Abe:2024tox}.\footnote{
Suppose the superpotential, $Y(\tau)Q\bar Q X$, where $Q$ and $\bar Q$ are hidden matter fields.
Then, we assume the condensation $\langle Q \bar Q \rangle = \Lambda^2$. That would lead to our superpotential term.}
For the K\"{a}hler invariant function $G=e^K |W_{\rm matter}|^2$ to be $\bar{\Gamma}$ invariant, the superpotential needs to carry  modular weight $-1$. Hence, the modular weights $k$ and $k_Y$ are related by
\begin{equation}
    -1 =k + k_Y.
\end{equation}
The Lagrangian is also symmetric under the CP by noting the following transformation properties of the modulus, the matter field, and modular forms
\cite{Novichkov:2019sqv}
\begin{align}
\label{eq: CP_trans_rules}
    \tau \xrightarrow{\rm CP} -\bar{\tau},\quad
    X \xrightarrow{\rm CP}  \bar{X}, \quad Y(\tau) \xrightarrow{\rm CP} e^{i \phi} \bar{Y}(\bar{\tau}),
\end{align}
where $\phi \in \mathbb{R}$ is a phase factor which depends on the normalization of $Y(\tau)$.
The superpotential then behaves as $W_{\rm matter} \xrightarrow{\rm CP} e^{i\phi} \bar{W}_{\rm matter}$ which indicates the CP symmetry. Therefore, the matter contributions are invariant under the extended modular symmetry $\bar{\Gamma}^*$.

\subsubsection{Supersymmetric vacua}
We study the supersymmetric vacuum satisfying the following conditions: 
\begin{equation}
    \partial_{\tau} W_{\rm matter} = \partial_X W_{\rm matter} = W_{\rm matter} =0.
\end{equation}
The solution is given by 
\begin{equation}
\label{eq: SUSY_Minimum_Matter}
    \langle X \rangle = 0,\quad \langle Y(\tau) \rangle = 0,
\end{equation}
provided $\langle \partial_{\tau} Y \rangle \neq 0$. This shows that $\tau$ is stabilized at a zero-point of $Y(\tau)$. 

\subsubsection{SUGRA scalar potential}
We compute the SUGRA scalar potential corresponding to the matter contributions to confirm the masses of fields at the potential minimum and their stability. We have the no-scale type scalar potential owing to $-2 \ln {\cal V}$ as assumed already:
\begin{equation}
   V = e^K K^{I\bar{J}} D_I W_{\rm matter} \overline{D_J W_{\rm matter}},
\end{equation}
where the indices are $I, J \in \{ \tau,\ X \}$. Within the region $|X| \ll 1$, the scalar potential $V$ can be expanded in powers of $X$ as
\begin{align}
    \begin{aligned}
V &= \frac{\Lambda^4}{{\mathcal{V}^2}}  {[2{\rm Im}(\tau)]^{k_Y}} |Y(\tau)|^2 \\
&\quad + \frac{\Lambda^4}{{\mathcal{V}^2}} {[2{\rm Im}(\tau)]^{-1}} \left[ 3|Y(\tau)|^2 + {|k_Y Y(\tau)+2i{\rm Im}(\tau)\partial_{\tau}Y(\tau)|^2} \right] |X|^2 + \mathcal{O}(|X|^4) \\
&=: V_0 + m_X^2 |X|^2 + \mathcal{O}(|X|^4).
    \end{aligned}
\end{align}
We find that the vacuum eq.~(\ref{eq: SUSY_Minimum_Matter}) is stable if $\langle \partial_{\tau} Y(\tau) \rangle \neq 0$. We can also check that the leading term $V_0$ and the mass term of $X$ are separately invariant under $\bar{\Gamma}^*$.

\subsubsection{Examples with $A_4$ modular forms}
\label{subsub: Model-I_A4_Ex}
\begin{figure}[H]
\centering
  \begin{minipage}[b]{0.46\linewidth}
    \centering
 \includegraphics[width=78mm]{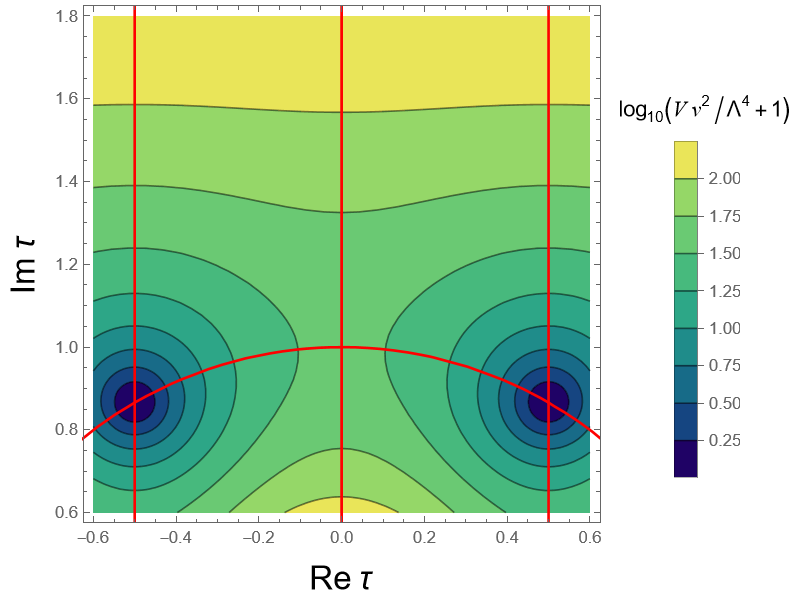}
\caption{Contour plot of 
$\log_{10}\left( {V_0 \mathcal{V}^2}/{\Lambda^4}+1 \right)$ when $A_4$ trivial singlet modular form $Y(\tau)=Y_{\bf 1}^{(4)}$ with weight $k_Y=4$ 
is chosen. 
The red curves denote the CP-conserving region, i.e. ${\rm Re}(\tau) = 0,\pm 1/2$ and $|\tau|=1$. This shows that the scalar potential $V_0=V|_{X=0}$ has its minimum at $\tau=\omega$.}
\label{fig: Potential_Weight4_Trivial_Singlet}
  \end{minipage} \
   \begin{minipage}[b]{0.46\linewidth}
    \centering
\includegraphics[width=78mm]{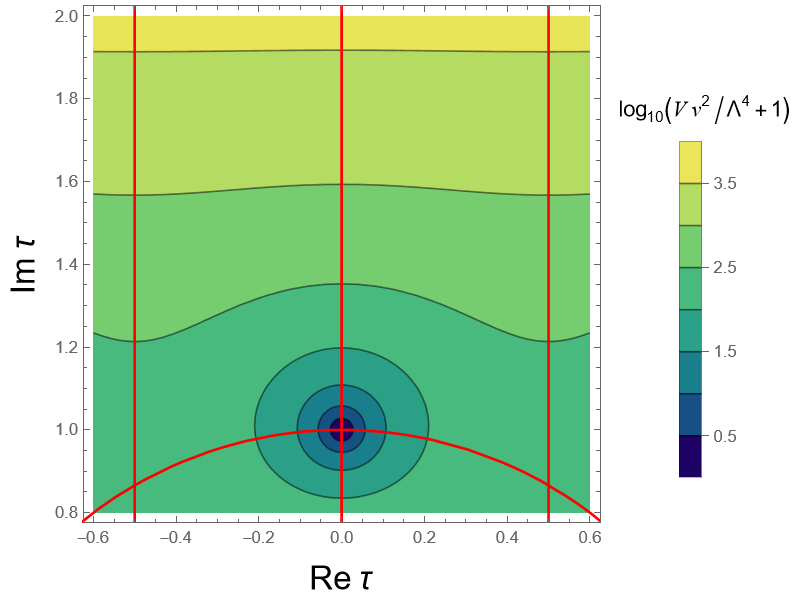}
\caption{Contour plot of 
$\log_{10}\left( {V_0 \mathcal{V}^2}/{\Lambda^4}+1 \right)$ when $A_4$ trivial singlet modular form 
$Y(\tau)=Y_{\bf 1}^{(6)}$
with weight $k_Y=6$ is chosen. The definition of this figure is similar to that of Fig.~\ref{fig: Potential_Weight4_Trivial_Singlet}.
This shows that the scalar potential $V_0=V|_{X=0}$ has its minimum at $\tau=i$.
\phantom{The red curves denote the CP-conserving region.}
}
\label{fig: Potential_Weight6_Trivial_Singlet}
  \end{minipage}
\end{figure}

As an illustrating example, let us choose
\begin{equation}
    Y(\tau) =  Y_{\bf{1}}^{(4)}(\tau),
 \end{equation}
where $Y_{\bf{1}}^{(4)}(\tau)$ denotes the $A_4$ trivial singlet modular form with weight $k_Y=4$. The definition of $Y_{\bf 1}^{(4)}$ is shown in Appendix \ref{appendix: modular_A4}. We have shown the contour plot of $\log_{10}(V_0 \mathcal{V}^2/\Lambda^4 + 1)$ in Fig.~\ref{fig: Potential_Weight4_Trivial_Singlet} where $V_0=V|_{X=0}$. 
In the figure, red curves correspond to the CP invariant lines, i.e. ${\rm Re}(\tau) = 0,\pm 1/2$ and $|\tau|=1$. It can be seen that the contour of the potential is perpendicular to the CP invariant lines, hence the general properties of the $\bar{\Gamma}^*$ invariant potential in Section~\ref{sec:general} are reflected. We find that the minimum of the potential is at the fixed point $\tau = \omega$.
One can check that $\tau = \omega$ is a zero-point of $Y_{\bf{1}}^{(4)}$ and its first derivative is nonzero,
\begin{equation}
    Y_{\bf{1}}^{(4)}(\omega) = 0,\quad \partial_{\tau}Y_{\bf{1}}^{(4)}\big|_{\tau = \omega} \simeq -6.04 i.
\end{equation}
Hence, $X$ is stabilized at $\langle X \rangle = 0$ and CP-conserving SUSY Minkowski vacuum is realized in this example.

Let us present one more example. We choose
\begin{equation}
    Y(\tau) = Y_{\bf{1}}^{(6)}(\tau),
 \end{equation}
where $Y_{\bf{1}}^{(6)}(\tau)$ denotes the $A_4$ trivial singlet modular form with weight $k_Y=6$.  The definition of $Y_{\bf 1}^{(6)}$ is shown in Appendix \ref{appendix: modular_A4}. 
In Fig.~\ref{fig: Potential_Weight6_Trivial_Singlet},
we have shown a similar plot of the potential consisting of $Y_{\bf{1}}^{(6)}(\tau)$.
The potential minimum is at $\tau = i$ which is a zero-point of the modular form $Y_{\bf 1}^{(6)}$ and its first derivative is nonzero,
\begin{equation}
\label{eq: A4_weight6_zero-point}
    Y_{\bf{1}}^{(6)}(i) = 0,\quad \partial_{\tau}Y_{\bf{1}}^{(6)}\big|_{\tau = i} \simeq -6.66i.
\end{equation}
Hence, $X$ is stabilized at $\langle X \rangle = 0$ and CP-conserving SUSY Minkowski vacuum is realized in this example.

We have observed that CP violation is hardly realized under the control of extended modular symmetry. It seems clear that some mechanisms to modify or break the modular symmetry will be useful in producing the CP violation.

\section{Spontaneous CP violation}

\label{sec:model}
We combine the contributions from the three-form fluxes and the matter to realize the spontaneous CP violation. The corresponding 4D $\mathcal{N}=1$ SUGRA Lagrangian is given by
\begin{align}
\begin{aligned}
    K &= K_{\rm moduli} + K_{\rm matter}, \\
    W &= W_{\rm flux} + W_{\rm matter},
\end{aligned}
\end{align}
where $K_{\rm moduli}$ and $W_{\rm flux}$ are shown in eqs.~(\ref{eq: Kahler_flux}) and (\ref{eq: simple_W_flux}), respectively. We study several concrete models by further specifying
$K_{\rm matter}$ and $W_{\rm matter}$. We will require modular invariance in the matter sector. This means that $K_{\rm matter}$ and $W_{\rm matter}$ respect the following discrete symmetry:
\begin{equation}
\label{eq: Full_Modular}
\left( \otimes_{i=1}^3
    \bar{\Gamma}_{\tau_i}  \otimes \bar{\Gamma}_S \right) \rtimes \mathbb{Z}_2^{\rm CP} ,
\end{equation}
where $\bar{\Gamma}_{\tau_i}$ and $\bar{\Gamma}_S$ correspond to the modular symmetry acting on the complex structure moduli $\tau_i$ and axio-dilaton $S$, respectively\footnote{
For instance, see Ref.~\cite{Yonekura:2024bvh} for duality group and references therein.
}. Note that we require $W_{\rm matter}$ to transform under CP as
\begin{equation}
\label{eq: W_matter_CP}
    W_{\rm matter} \xrightarrow{\rm CP} - \bar{W}_{\rm matter}.
\end{equation}
This is necessary for the total superpotential $W$ to transform as
\begin{equation}
    W \xrightarrow{\rm CP} - \bar{W}.
\end{equation}
Note that $W_{\rm flux}$ in eq.~(\ref{eq: simple_W_flux}) transforms as $W_{\rm flux} \xrightarrow{\rm CP} - \bar{W}_{\rm flux}$.

\subsection{Model I}
Here, we study the simplest model, introducing a single matter field $X$ carrying modular weights $k \in \mathbb{Z}$ under $\bar{\Gamma}_{\tau_1}$ and $-1$ under $\bar{\Gamma}_{\tau_2}, \bar{\Gamma}_{\tau_3}, \bar{\Gamma}_{S}$.
The matter K\"{a}hler potential is given by 
\begin{align}
    \begin{aligned}
    K_{\rm matter}  &=   Z|X|^2,
    \end{aligned}
\end{align}
and the matter superpotential is given by 
\begin{equation}
\label{eq: ModelI_W_matter}
    W_{\rm matter} = \Lambda^2 Y(\tau_1) X,
\end{equation}
where
\begin{equation}
     Z =(-i\tau_1 +i\bar{\tau}_1)^k (-i\tau_2 +i\bar{\tau}_2)^{-1}(-i\tau_3 +i\bar{\tau}_3)^{-1}(-iS +i\bar{S})^{-1},
\end{equation}
to preserve the discrete symmetry eq.~(\ref{eq: Full_Modular}). 
We require that the trivial singlet modular form
$Y(\tau_1)$ carries a modular weight $k_Y = -(1+k)$ under $\bar{\Gamma}_{\tau_1}$. Furthermore, we choose 
an appropriate normalization of $Y(\tau_1)$ to realize the following CP transformation
\begin{equation}
\label{eq: CP_Y_negative}
    Y(\tau_1) \xrightarrow{\rm CP} - \bar{Y}(\bar{\tau}_1).    
\end{equation}
This corresponds to $\phi = \pi$ in eq.~(\ref{eq: CP_trans_rules}).
Then, the matter superpotential (\ref{eq: ModelI_W_matter}) behaves as in eq.~(\ref{eq: W_matter_CP}). 

We solve the SUSY conditions:
\begin{equation}
    \partial_{\tau_i} W = 0,\quad \partial_S W = 0,\quad  \partial_X W = 0,\quad W=0.
\end{equation}
The solution is given by 
\begin{align}
\begin{aligned}
\tau_1 \tau_2 = \frac{b_3}{a^0},\quad \tau_3 = fS,\quad Y(\tau_1)=0,\quad X=0\ ({\rm if\ }\partial_{\tau_1}Y \neq 0). 
\end{aligned}
\end{align}
We find $\tau_1$ is stabilized at a zero-point of the modular forms $Y(\tau_1)$ causing the lifting of the flat direction $\tau_1 \tau_2 = b_3/a^0$. As a result, simultaneous stabilization of $\tau_2$ occurs. On the other hand, there remains a flat direction $\tau_3 = f S$.

\subsubsection{Examples with $A_4$ modular form}
\label{subsubsection: Examples_CPV_ModelI}
If we assign some specific flux quanta, we can realize the spontaneous CP violation by the VEV of $\tau_2$. 

\paragraph{Weight 4 trivial singlet:}
As an illustrating example, let us choose
\begin{equation}
    Y(\tau_1) =i~Y_{\bf{1}}^{(4)}(\tau_1),
\end{equation}
where $Y_{\bf{1}}^{(4)}(\tau_1)$ is a trivial singlet $A_4$ modular form with weight $k_Y=4$.
It is defined in Appendix \ref{appendix: modular_A4}. Since the normalization of $Y_{\bf{1}}^{(4)}$ 
corresponds to $\phi =0$ in eq.~(\ref{eq: CP_trans_rules}), we have introduced the overall factor $i$ 
for $Y(\tau_1)$ to transform as in eq.~(\ref{eq: CP_Y_negative}).
The complex structure modulus is stabilized at $\tau_1 = \omega$ which is a zero-point of $Y_{\bf 1}^{(4)}$. 
Then, this leads to the simultaneous stabilization of $\tau_2$ at
\begin{equation}
\label{eq: VEV_tau2}
\tau_2 = \frac{b_3}{a^0} \frac{1}{\omega} . 
\end{equation}

We present some examples of flux quanta that realize the spontaneous CP violation. 
Let us choose
\begin{equation}
\label{eq: CPV_Ex1}
    a^0 =2 ,\quad b_3 = -3 ,\quad  f=1,
\end{equation}
satisfying the tadpole cancellation condition, $n_{\rm flux} = 12 \leq 32$. Then $\tau_2$ is stabilized at 
\begin{equation}
\label{eq: Model_I_eg1}
    \tau_2 = -\frac{3}{2 \omega} = \frac{3}{4} + \frac{3\sqrt{3}}{4}i,
\end{equation}
where CP is violated. 
Next, let us choose
\begin{equation}
\label{eq: CPV_Ex2}
    a^0 =3,\quad b_3 = -4 ,\quad  f=1,
\end{equation}
satisfying the tadpole cancellation condition, $n_{\rm flux} = 24 \leq 32$. Then $\tau_2$ is stabilized at 
\begin{equation}
\label{eq: Model_I_eg2}
   \tau_2 = -\frac{4}{3 \omega} = \frac{2}{3} + \frac{2i}{\sqrt{3}},
\end{equation}
where CP is violated.

\paragraph{Weight 6 trivial singlet:}
As an illustrating example, let us choose
\begin{equation}
    Y(\tau_1) =i~ Y_{\bf{1}}^{(6)}(\tau_1),
\end{equation}
where $Y_{\bf{1}}^{(6)}(\tau_1)$ is a trivial singlet $A_4$ modular form with weight $k_Y=6$.
It is defined in Appendix \ref{appendix: modular_A4}. Since the normalization of $Y_{\bf{1}}^{(6)}$ corresponds to $\phi =0$ in eq.~(\ref{eq: CP_trans_rules}), we have introduced the overall factor $i$
for $Y(\tau_1)$ to transform as in eq.~(\ref{eq: CP_Y_negative}).
As in eq.~(\ref{eq: A4_weight6_zero-point}), $Y_{\bf 1}^{(6)}$ has a zero-point at $\tau_1 =i$, hence stabilization to $\langle \tau_1 \rangle=i$ is possible. In this case, there is no CP violation because we find ${\rm Re}(\tau_2) =0$ regardless of the flux quanta assignment. On the other hand, if $\tau_1$ stabilizes at different zero-points of $Y_{\bf 1}^{(6)}$ such as $\tau_1 = i \xrightarrow{T} 1+i$, we can realize spontaneous CP violation.
Let us focus on the case $\langle \tau_1 \rangle=1+i$ and assign following flux quanta: 
\begin{equation}
    a^0 =2,\quad b_3 = -5 ,\quad  f=1.
\end{equation}
The tadpole cancellation condition is satisfied, $n_{\rm flux} = 20 \leq 32$. Then $\tau_2$ is stabilized at 
\begin{equation}
\label{eq: Model_I_eg3}
    \tau_2 =\frac{-5}{2(1+i)} = \frac{5}{4}(-1+i),
\end{equation}
where CP is violated. In short, CP can be spontaneously broken depending on the VEV of $\tau_1$ and the choice of flux quanta.

\subsubsection{Discrete symmetry}
\label{subsubsection: Discrete_SYM_Model_I}
Here, we discuss the discrete symmetries behind the spontaneous CP violation caused by the VEV of $\tau_2$. Under $\langle X \rangle = 0$, the scalar potential $V$ is reduced to  
\begin{equation}
\label{eq: Model_I_scalar_potential}
    V\big|_{X=0} = \frac{\Lambda^4}{\mathcal{V}^2} [2{\rm Im}(\tau_1)]^{k_Y}|Y_{\bf 1}^{(k_Y)}(\tau_1)|^2 + \frac{2 |\tau_3-fS|^2\  |a^0 \bar{\tau}_1 \tau_2 - b_3|^2 \ +\  2  |\tau_3 - f \bar{S}|^2\  |a^0 \tau_1 \tau_2  - b_3|^2}{\mathcal{V}^2 [2{\rm Im}(\tau_1)][2{\rm Im}(\tau_2)][2{\rm Im}(\tau_3)][2{\rm Im}(S)]}.
\end{equation}
The first term corresponds to matter contributions. The remaining terms correspond to the three-form fluxes. Restricting to the flat directions generated by the fluxes, the potential is further reduced to 
\begin{equation}
    V \rightarrow V_{\rm matter} = \frac{\Lambda^4}{\mathcal{V}^2} [2{\rm Im}(\tau_1)]^{k_Y}|Y_{\bf 1}^{(k_Y)}(\tau_1)|^2.
\end{equation}
This is symmetric under the extended modular group $SL(2,\mathbb{Z})_{\tau_1} \rtimes \mathbb{Z}_2^{\rm CP}$, implying that the VEV of $\tau_1$ hardly realizes CP violation. On the other hand, rewriting $V_{\rm matter}$ as a function of $\tau_2$ and $\bar{\tau}_2$ with $\tau_1 = b_3/(a^0 \tau_2)$, we find
\begin{equation}
    V_{\rm matter} = \frac{\Lambda^4}{\mathcal{V}^2} \left[ \frac{|a^0|}{|b_3|}  {2{\rm Im}(\tau_2)} \right]^{k_Y} \left|Y_{\bf 1}^{(k_Y)}\left( \frac{|a^0|}{|b_3|} \tau_2 \right)\right|^2,
\end{equation}
where we have used the invariance of $V_{\rm matter}$ under $S \in SL(2,\mathbb{Z})_{\tau_1}$ acting as $\tau_1 \to -1/\tau_1 = (|a^0|/|b_3|) \tau_2$.
This is CP invariant, however no longer symmetric under $SL(2,\mathbb{Z})_{\tau_2}$ in general. Thus, we may have a chance to obtain CP violation by $\tau_2$.
One can check that the discrete symmetry of $V_{\rm matter}$ is given by 
\begin{equation}
\label{eq: scale_SL2Z}
    \gamma \in \left\{ 
    \begin{pmatrix}
        a &  b \\
        c & d 
    \end{pmatrix} \in SL(2,\mathbb{Q}) \Bigg| \begin{pmatrix}
        a &  b  \\
        c & d 
    \end{pmatrix} \equiv
    \begin{pmatrix}
      *  
      & 0 \pmod{|b_3/a^0|} \\
        0  \pmod{|a^0/b_3|}  & *
    \end{pmatrix} \right\},
\end{equation}
where $*$ denotes unspecified integers and $\gamma$ acts
on $\tau_2$ in a familiar way,
\begin{equation}
    \tau_2 \xrightarrow{\gamma} \frac{a\tau_2+b}{c \tau_2 + d}.
\end{equation}
One can check that the set of matrices in eq.~(\ref{eq: scale_SL2Z}) form a group isomorphic to $SL(2,\mathbb{Z})_{\tau_2}$.
This can be understood as follows. We define a matrix $P$ by
\begin{equation}
    P = \begin{pmatrix}
     \sqrt{|a^0|/|b_3|} & 0 \\
     0 &  \sqrt{|b_3|/|a^0|}
    \end{pmatrix}.
\end{equation}
Then, one can verify that the following map 
\begin{equation}
  \gamma \rightarrow  P \gamma P^{-1} \in SL(2,\mathbb{Z})_{\tau_2}, 
\end{equation}
from (\ref{eq: scale_SL2Z}) to $SL(2,\mathbb{Z})_{\tau_2}$ is an isomorphism.
Note that if $|b_3|/|a^0|=1$, 
the group (\ref{eq: scale_SL2Z})
is identical to $SL(2,\mathbb{Z})_{\tau_2}$.
We notice that the discrete group contains the following two elements analogous to $T$ and $S$ transformations of $SL(2,\mathbb{Z})$: 
\begin{align}
       \hat{T} = 
    \begin{pmatrix}
    1 & |b_3|/|a^0| \\ 0 & 1    
    \end{pmatrix}, \quad
    \hat{S}=\begin{pmatrix}
        0 & |b_3|/|a^0| \\ - |a^0|/|b_3| & 0 
    \end{pmatrix}.
\end{align}
Symmetries of the scalar potential $V_{\rm matter}$ under $\hat{T}$ and $\hat{S}$ transformations imply the following properties:
 \begin{align}
 \label{eq: hat_T_V}
     \frac{\partial V_{\rm matter}}{\partial s}(s,t) = 0, \quad  s\equiv 0 \pmod{|b_3/(2a^0)|},
\end{align}
where $\tau_2 = s + i t$, and 
 \begin{equation}
 \label{eq: hat_S_V}
         \frac{\partial V_{\rm matter}}{\partial r}(r, \theta) = 0,\quad  r = \frac{|b_3|}{|a^0|},
 \end{equation} 
where $\tau_2 = r e^{i \theta}$. Recalling the examples with $A_4$ modular forms in sec.~\ref{subsubsection: Examples_CPV_ModelI}, we find that the CP-violating VEVs of $\tau_2$ shown by eqs.~(\ref{eq: Model_I_eg1}), (\ref{eq: Model_I_eg2}) and (\ref{eq: Model_I_eg3}) are on the line 
${\rm Re}(\tau_2) = \pm |b_3/(2a^0)|$ 
as predicted by eq.~(\ref{eq: hat_T_V}). Furthermore, (\ref{eq: Model_I_eg1}) and (\ref{eq: Model_I_eg2}) are also on the curve $|\tau_2| = |b_3|/|a^0|$ as predicted by eq.~(\ref{eq: hat_S_V}). In fact, (\ref{eq: Model_I_eg1}) and (\ref{eq: Model_I_eg2}) correspond to the fixed point under $\hat{T} \hat{S}$. Similarly, (\ref{eq: Model_I_eg3}) corresponds to the fixed point under $\hat{S} \hat{T} \hat{S} (\hat{S} \hat{T})^{-1}$.
One could simply understand these results by noticing that the modulus space of $\tau_2$ is obtained via the scale transformation:
\begin{equation}
\label{eq: tau_rescale}
    \tau_2 = \frac{|b_3|}{|a^0|} \tau,
\end{equation}
where $\tau$ corresponds to the modulus on which $SL(2,\mathbb{Z})$ acts. Therefore, the spontaneous CP violation is readily induced if the VEV of $\tau_2$ breaks CP even when the value of $\tau$ is CP-conserving, typically the fixed points.

We have observed that there are preferred flux quanta for the realization of spontaneous CP violation. From eq.~(\ref{eq: hat_T_V}) we can see that when the flux quanta satisfy 
\begin{equation}
\label{eq: CPV_condition}
    b_3 \nmid a^0 \quad  {\rm and}\quad   a^0 \nmid b_3,
\end{equation}
CP violation is easily realized, where $b_3 \nmid a^0$ denotes $a^0$ is not divisible by $b_3$. This is because the lines ${\rm Re}(\tau_2) = m |b_3/(2a^0)|,\ (m\in \mathbb{Z})$ do not coincide with the CP-conserving lines ${\rm Re}(\tau_2) \equiv 0 \pmod{1/2}$, in general. The condition (\ref{eq: CPV_condition}) is satisfied in all the CP-violating examples in sec.~\ref{subsubsection: Examples_CPV_ModelI}.
On the other hand, when $b_3/a^0 = n \in \mathbb{Z}$, CP is hardly violated. From eq.~(\ref{eq: hat_T_V}), $\tau_2$ tends to stabilize on the line ${\rm Re}(\tau_2) = {n}/{2}$ where CP is conserved. The case that $b_3/a^0 = -1/n~(n \in \mathbb{Z})$ is similar by noting that $\tau_2^{(S)} = -1/\tau_2 = n \tau_1$. If $\tau_1$ is stabilized at ${\rm Re}(\tau_1) \equiv 0 \pmod{1/2}$, the VEV of $\tau_2$ will not break CP.\footnote{If we consider a very special set-up, CP violation may be realized even when eq.~(\ref{eq: CPV_condition}) is not satisfied. For example, choosing $|b_3/a^0|=1/7$ and $\tau_1 = ST^2i= -0.4 + 0.2i$, we obtain $\tau_2 \simeq 0.286 +0.143i$ which cannot be mapped to a CP-conserving point by a modular transformation.}

We have found that the condition eq.~(\ref{eq: CPV_condition}) is satisfied under the tadpole cancellation condition (\ref{eq: tadpole2}) only if we introduce odd-valued flux quanta. We give proof in Appendix \ref{appendix: CP_condition}. This means that introducing exotic $O3'$ planes increases the likelihood of spontaneous CP violation in Model I. 
If we relax the tadpole cancellation condition, all the flux quanta $a^0, b_3, c^3$ and $d_0$ can be even.
For example, the condition~(\ref{eq: CPV_condition}) can be 
satisfied for $(a^0,b_3,c^3,d_0)=(4,-6,-4,-6)$ with $f=1$, although they lead to $n_{\rm flux}=48$.


In Model I, we have succeeded in realizing the spontaneous CP violation by the VEV of $\tau_2$. The obtained potential minima correspond to the SUSY Minkowski vacua.
However, $\tau_3$ and  $S$ as well as $\mathcal{V}$ are not stabilized.
In the next subsection, we will stabilize  
$\tau_3$ and $S$.

\subsection{Model II}
Here, we consider an extension of Model I by introducing one additional matter field $x$ carrying modular weights $l \in \mathbb{Z}$ under $\bar{\Gamma}_{\tau_3}$ 
and $-1$ under $\bar{\Gamma}_{\tau_1}, \bar{\Gamma}_{\tau_2}, \bar{\Gamma}_{S}$. The matter K\"{a}hler potential is given by 
\begin{equation}
    K_{\rm matter}  =  Z|X|^2 + z|x|^2,
\end{equation}
and the matter superpotential is given by 
\begin{equation}
    W_{\rm matter} = \Lambda^2 ( Y(\tau_1) X + \alpha \ y(\tau_3) x),\quad \alpha \in \mathbb{R},
\end{equation}
where
\begin{align}
\begin{aligned}
Z &=(-i\tau_1 +i\bar{\tau}_1)^k(-i\tau_2 +i\bar{\tau}_2)^{-1}(-i\tau_3 +i\bar{\tau}_3)^{-1}(-iS +i\bar{S})^{-1},
\\
z&=(-i\tau_1 +i\bar{\tau}_1)^{-1}(-i\tau_2 +i\bar{\tau}_2)^{-1}(-i\tau_3 + i\bar{\tau}_3)^l(-iS +i\bar{S})^{-1},
\end{aligned}
\end{align}
to preserve the discrete symmetry eq.~(\ref{eq: Full_Modular}). 
We require that the trivial singlet modular forms
$Y(\tau_1)$ and $y(\tau_3)$ carry modular weights $k_Y = -(1+k)$ under $\bar{\Gamma}_{\tau_1}$ and $k_y = -(1+l)$ under $\bar{\Gamma}_{\tau_3}$, respectively. We emphasize that the reality of the dimensionless constant $\alpha$ is required for CP invariance \cite{Novichkov:2019sqv}.\footnote{
The results do not change even for $\alpha \in \mathbb{C}$ since the $\alpha$ contributes to the scalar potential as $|\alpha|^2$.
}

We solve the SUSY conditions:
\begin{equation}
    \partial_{\tau_i} W = 0,\quad \partial_S W = 0,\quad  \partial_X W = 0,\quad \partial_xW=0,\quad W=0.
\end{equation}
The solution is given by 
\begin{align}
\begin{aligned}
\label{eq: Model_II_Solution}
 \tau_1 \tau_2 = \frac{b_3}{a^0},\quad \tau_3 = fS,\quad Y(\tau_1)=0,\quad y(\tau_3) =0,\\
        \quad X=0\ ({\rm if\ }\partial_{\tau_1}Y \neq 0),\quad x=0\ ({\rm if\ }\partial_{\tau_3}y \neq0).
\end{aligned}
\end{align}
We find that $\tau_1$ and $\tau_3$ are stabilized at zero-points of the modular forms $Y(\tau_1)$ and $y(\tau_3)$, respectively. As a result, all the flat directions are lifted and $\tau_2$ and $S$ are stabilized accordingly. 

\subsubsection{Spontaneous CP violation in the weak coupling regime}
As an illustrating example, we choose
\begin{equation}
\label{eq:modular-modelII}
    Y(\tau_1) = i~Y_{\bf{1}}^{(4)}(\tau_1),\quad  y(\tau_3) = i~Y_{\bf{1}}^{(6)}(\tau_3),
 \end{equation}
where $Y_{\bf{1}}^{(4)}$ and $Y_{\bf{1}}^{(6)}$ denote the $A_4$ trivial singlets of weight $4$ and $6$, respectively. The overall factors $i$ 
are introduced to satisfy eq.~(\ref{eq: W_matter_CP}).
We can realize the spontaneous CP violation under the weak coupling regime ${\rm Im}(S) > 1$ if we choose specific flux quanta.
For example, let us take
\begin{equation}
    a^0 =4 ,\quad b_3 = -6 ,\quad  f=\frac{1}{2},
\end{equation}
satisfying the tadpole cancellation condition, $n_{\rm flux} = 24 \leq 32$. Note that all the flux quanta are integer-valued. The non-vanishing quanta other than $a^0, b_3$
are given by
\begin{equation}
    c^3 = -2,\quad d_0=-3,
\end{equation}
which follows from eq.~(\ref{eq: Restricted_Flux_Quanta}).
Consequently, the complex structure moduli and the axio-dilaton are stabilized at
\begin{equation}
\label{eq: Model_II_A4}
    \tau_1 = \omega,\quad 
    \tau_2 = \frac{3}{4} + \frac{3\sqrt{3}}{4}i,
    \quad \tau_3 = i,\quad S = 2i.
\end{equation}
We find that CP is spontaneously broken by the VEV of $\tau_2$ under the weak coupling regime, i.e. ${\rm Im}(S) > 1$.
If we relax the tadpole cancellation, we can realize the identical CP-violating vacuum only using even-valued flux quanta. For example, choosing $(a^0,b_3,c^0,d_0)=(8,-12,-4,-6)$ with $f=1/2$ recovers the VEVs in eq.~(\ref{eq: Model_II_A4}), although $n_{\rm flux} = 96$.

We have focused on the odd polynomial form of the moduli superpotential eq.~(\ref{eq: W_flux_CP_invariant}) in Models I and II.
Similarly, we can discuss the even polynomial form of the moduli superpotential eq.~(\ref{eq: W_even}).
For the even polynomial form, we replace eq.~(\ref{eq:modular-modelII}) by $Y(\tau_1) = Y_{\bf{1}}^{(4)}(\tau_1)$ and $  y(\tau_3) =Y_{\bf{1}}^{(6)}(\tau_3)$, which 
correspond to $\gamma=\phi=0$, in Model II, and we do a similar replacement in Model I.

\subsubsection{Moduli masses}
Since the obtained vacuum preserves SUSY, we compute the supersymmetric mass matrix defined by $m_{AB}=\partial_A \partial_B W$.
The non-vanishing components are given by
\begin{align}
\begin{aligned}
    m_{\tau_1 \tau_3} &= m_{\tau_3 \tau_1} = a^0 \langle \tau_2 \rangle, \\
    m_{\tau_1 S} &= m_{S \tau_1} = - f a^0 \langle \tau_2 \rangle, \\
    m_{\tau_1 X} &=  m_{X \tau_1} = \Lambda^2 \langle \partial_{\tau_1}Y(\tau_1) \rangle, \\
    m_{\tau_2 \tau_3} &= m_{\tau_3 \tau_2} = a^0 \langle \tau_1 \rangle, \\
    m_{\tau_2 S} &= m_{S \tau_2} = - f a^0 \langle \tau_1 \rangle, \\
    m_{\tau_3 x} &=  m_{x \tau_3} = \alpha^2 \Lambda^2  \langle 
    \partial_{\tau_3} y(\tau_3) \rangle.
\end{aligned}
\end{align}
The determinant of $m_{AB}$ is 
\begin{equation}
\label{eq: det_supersymmetric_mass}
    \det{(m_{AB})} =  - (a^0)^2 f^2 \alpha^4 \Lambda^8 \langle \tau_1 \rangle^2  \langle \partial_{\tau_1} Y(\tau_1) \rangle^2  \langle \partial_{\tau_3} y(\tau_3) \rangle^2,
\end{equation}
which is generally non-zero unless $\alpha =0$. This assures that $\tau_i, S, X, x$ are all stabilized. 
Let us comment on the masses of the fields. 
Due to the three-form fluxes, only half of the moduli fields acquire masses of order $\mathcal{O}(M_{\rm Pl})$ as shown in eq.~(\ref{eq: Mass_scale_Flux}). The remaining half of the moduli fields are stabilized by the matter contributions giving masses of order $\mathcal{O}(\Lambda^2/M_{\rm Pl})$ if $\alpha \sim \mathcal{O}(1)$. Masses of the matter fields $X, x$ are also around $\mathcal{O}(\Lambda^2/M_{\rm Pl})$ provided $\langle \partial_{\tau_1}Y(\tau_1) \rangle \neq 0,\langle \partial_{\tau_3}y(\tau_3) \rangle \neq 0$.


We have realized the spontaneous CP violation in the weak coupling regime ${\rm Im}(S) > 1$ by an appropriate choice of flux quanta. All the complex structure moduli and the axio-dilaton are stabilized. The obtained potential minima correspond to the SUSY Minkowski vacua.
Note that the K\"{a}hler modulus is still unfixed.
In the next subsection, we will show the K\"{a}hler modulus can be 
stabilized without disturbing our results.

\subsection{K\"{a}hler moduli stabilization}
\label{sec:results}
We have shown the model, where all of the moduli are stabilized except the  K\"{a}hler modulus 
and the CP is violated spontaneously.
In this section, we study the stabilization of 
the K\"{a}hler modulus $T$ which is related to the volume as $\mathcal{V} \simeq [2 {\rm Im}(T)]^{3/2}$.
For illustration of the K\"{a}hler modulus stabilization,
we introduce non-perturbative effects to the superpotential: 
\begin{equation}
\label{eq: W_fmn}
    W = W_{\rm flux} + W_{\rm matter}  + W_{\rm np},
\end{equation}
where
\begin{equation}
    W_{\rm np} = \sum_m C_m e^{ia_m T + i b_m S}.
\end{equation}
Possible origins of $W_{\rm np}$ are the D-brane instanton effects with $a_m, b_m = 2\pi$ or gaugino condensation of the gauge theory on $D7$ branes. 
We discuss the  K\"{a}hler modulus stabilization 
based on Model II, where all fields except for the K\"{a}hler modulus are stabilized in a SUSY Minkowski vacuum.
Following Ref.~\cite{Ishiguro:2022pde},
we consider that the constant term in the effective superpotential is generated by the axio-dilaton dependent non-perturbative effects:
\begin{equation}
    W_{\rm eff} = W_{\rm np}(\langle S \rangle, T),
\end{equation}
where we presume $\langle S \rangle \sim 2i$ as a result of three-form fluxes and the matter contributions, cf. eq.~(\ref{eq: Model_II_A4}).
One obtains the following KKLT-type superpotential \cite{Kachru:2003aw}:
\begin{equation}
    W_{\rm eff} \simeq \langle {\Lambda'}^3 e^{ibS} \rangle + C e^{iaT},
\end{equation}
where $\Lambda'$ is a parameter with mass dimension one. As a result, the K\"{a}hler modulus is stabilized at $T=\langle T \rangle$ satisfying
\begin{equation}
\label{eq: DT=0}
    D_T W_{\rm eff} = \partial_T W_{\rm eff} + (\partial_T K) W_{\rm eff}=0,
\end{equation}
which approximately gives us
\begin{equation}
a\ {\rm Im} \langle  T \rangle \simeq  \ln{(C/w_0)},
\end{equation}
where $|w_0|= | \langle {\Lambda'}^3 e^{ibS} \rangle | \ll 1~(=M_{\rm Pl}^3)$.
Here and hereafter, we assume that $C$ is a constant, particularly $|C|=1$. One then finds, ${\rm Im} \langle T \rangle \sim 2 > 1$ when we choose $\Lambda' = 1, a=b=2 \pi $.


In the previous discussion, we have assumed that the VEVs of
the complex structure moduli, axio-dilaton, and matter fields are given by eq.~(\ref{eq: Model_II_Solution}) in Model II where contributions from the three-form flux and the moduli-matter couplings are considered without the non-perturbative effects. Then, we have separately analyzed the dynamics of the K\"{a}hler modulus. This is an approximation. Thus, there will be deviations in the VEVs of the fields if we have simultaneously considered the non-perturbative effects along with the fluxes and the matter effects. 
Suppose that
true vacuum satisfies
\begin{equation}
\label{eq: True_SUSY}
    D_I W = 0,
\end{equation}
where $W$ is given by eq.~(\ref{eq: W_fmn}).
Following Refs.~\cite{Abe:2006xp,Abe:2007yb}, we estimate the deviations.
We write the true vacuum values up to a linear order 
of deviations as 
\begin{align}
\begin{aligned}
\label{eq: deviations}
\tau_i &= \langle \tau_i \rangle + \delta \tau_i, \\
S &= \langle S \rangle + \delta S, \\
X &= \langle X \rangle + \delta X = \delta X, \\
x &= \langle x \rangle + \delta x = \delta x, \\
T &= \langle T \rangle + \delta T,
\end{aligned}
\end{align}
where $\langle \tau_i \rangle, \langle S \rangle, \langle X \rangle$ and $\langle x \rangle$ are the VEVs shown in eq.~(\ref{eq: Model_II_Solution}). The VEV $\langle T \rangle$ is determined by eq.~(\ref{eq: DT=0}). 
These values correspond to the reference point to 
evaluate the true vacuum \cite{Abe:2006xp,Abe:2007yb}.

We consider the  K\"{a}hler invariant quantity $G=K+\ln{|W|^2}$ and analyze $G_A = \partial_A G = 0$ which is equivalent to eq.~(\ref{eq: True_SUSY}). The index $A$ denotes both the holomorphic and anti-holomorphic fields:
\begin{equation}
\label{eq: Set_of_fields}
A = \{\tau_i,S,X,x,T, \bar{\tau}_i, \bar{S}, \bar{X}, \bar{x}, \bar{T} \}.
\end{equation}
From eq.~(\ref{eq: deviations}), $G_A$ is expanded as
\begin{equation}
G_A = G_A |_{\langle \rangle} + \delta \phi^B G_{AB}|_{\langle \rangle} + \mathcal{O}((\delta \phi)^2),
\end{equation}
where $\phi$ denotes the fields in eq.~(\ref{eq: Set_of_fields}). The symbol $|_{\rm \langle \rangle}$ indicates that we take values at the reference point, i.e. $\phi = \langle \phi \rangle$. Under the assumption $a,b > 1$, one finds that non-vanishing components of $G_{AB}$ satisfy
\begin{equation}
    G_{IJ}, G_{\bar{I}\bar{J}} \gg G_{\bar{I} J}, G_{I\bar{J}}.
\end{equation}
This allows us to diagonalize the holomorphic and anti-holomorphic parts of
$G_{AB}$ and $G^{AB}$ separately in the leading approximation, where $G^{AB}$ is the inverse matrix of the $G_{AB}$.
Thus, we obtain
\begin{equation}
    \delta \phi^I = - G^{IJ}|_{\langle \rangle}  G_J|_{\langle \rangle}  + (\mathcal{O}(\delta \phi)^2).
\end{equation}
By explicit computation, we find
\begin{align}
\begin{aligned}
\delta \tau_1 &= \mathcal{O}(\varepsilon^2), \\
\delta \tau_2 &=  - \frac{W_{\rm eff}}{W_{\tau_2 S}} 
 G_S \Big|_{\langle \rangle} +   \mathcal{O}(\varepsilon^2),\\
\delta \tau_3 &= \mathcal{O}(\varepsilon^2), \\
\delta S &= - \frac{W_{\rm eff}}{W_{\tau_2 S}} G_{\tau_2} \Big|_{\langle \rangle} + \mathcal{O}(\varepsilon^2),  \\
\delta X &= \frac{W_{\rm eff}}{W_{\tau_1 X}} \left( \frac{W_{\tau_1 S}}{W_{\tau_2 S} } G_{\tau_2} - G_{\tau_1}
\right) \bigg|_{\langle \rangle} + \mathcal{O}(\varepsilon^2) =
- 2 \frac{W_{\rm eff}}{W_{\tau_1 X}} G_{\tau_1} \Big|_{\langle \rangle} + \mathcal{O}(\varepsilon^2), \\
\delta x &= \frac{W_{\rm eff}}{W_{\tau_3 x}} \left( \frac{W_{\tau_2 \tau_3}}{W_{\tau_2 S} } G_{S} - G_{\tau_3}
\right) \bigg|_{\langle \rangle} + \mathcal{O}(\varepsilon^2)
, \\
\delta T &= -\frac{G_{ST}}{G_{TT}} \bigg|_{\langle \rangle} \delta S ,
\end{aligned}    
\end{align}
where $\varepsilon = |W_{\rm eff}| / (\Lambda^2 M_{\rm Pl}) \ll 1$ is assumed. This shows that deviations from the reference point are exponentially suppressed.

The vacuum energy is negative, 
$V=-3\langle e^K |W_{\rm eff}|^2 \rangle <0$.
We need to uplift the vacuum energy by SUSY breaking due to non-vanishing F-terms, D-terms and/or anti-branes.
Required F-terms and D-terms are small enough compared with 
the above moduli masses and matter mass.
Thus, deviations of those VEVs due to uplifting would be 
smaller than the above shift.

In Appendix B, we give another model leading to 
the spontaneous CP violation, where the de Sitter vacuum is realized.

Before ending this section, we comment on 
the stabilized values,  $S=2i$.
This value is interesting phenomenologically.
Suppose that the gauge kinetic function of the standard gauge group 
$SU(3) \times SU(2)\times U(1)_Y$ is given by $f_h=-iS$, where $f_h$ is a (unified) holomorphic gauge coupling in a supersymmetric standard model at the UV cutoff scale.
Then, we find $\alpha^{-1} = 8 \pi \approx 25$, where 
$\alpha$ denotes the fine structure constant.
This value is almost the unified gauge coupling of the minimal supersymmetric 
standard model at the GUT scale.
Furthermore, the value ${\rm Re}S=0$ leads to the vanishing QCD phase $\theta_0=0$ 
at tree level.
We may have corrections.
Indeed, we have added the correction term, $\Delta W = W_{\rm np}$.
That can shift ${\rm Re}S$, which is estimated by 
\begin{align}
    \delta S\sim - i 
    \frac{\Delta W}{W_{\tau_2 S}}
    \sim ~ - i \frac{m_{3/2}}{M_{\rm Pl}}
    e^{i \varphi}, 
\end{align}
where we write
$\langle \Delta W \rangle  e^{-i\arg{\langle W_{\tau_2 S} \rangle}} = m_{3/2}M_{\rm Pl}^2 e^{i \varphi}$ and 
$m_{3/2}$ denotes the (real) gravitino mass.
That is, we find 
\begin{align}
  |  {\rm Re} (\delta S) | \approx |{\sin \varphi}| ~ \frac{m_{3/2}}{M_{\rm Pl}}.
\end{align}
If 
$m_{3/2} |{\sin \varphi}|/M_{\rm Pl} < 10^{-10}$
, this shift ${\rm Re} (\delta S)$ is 
consistent with current experiments of electric dipole moment of neutrons.
That gives us the upper bound for the gravitino mass, 
$m_{3/2} < 10^{-10}\times M_{\rm Pl}/|\sin \varphi|$.
We may have other corrections. 
If such corrections 
are similar to or smaller than $ m_{3/2}M_{\rm Pl}^2 e^{i \varphi}$, 
${\rm Re}(\delta S)$ would be still consistent with experiments. 
However, the true QCD phase $\theta$ has the corrections due to quark masses $M_q$,
\begin{align}
    \theta = \theta_0 + {\rm arg}({\rm det} M_q).
\end{align}
Recently, the mass matrices leading to ${\rm arg}({\rm det} M_q)=0$ were 
studied \cite{Feruglio:2023uof,Petcov:2024vph,Penedo:2024gtb}.\footnote{
There are other approaches relevant to the modular symmetry including the axion \cite{Ibanez:1991qh,Kobayashi:2020oji,Ahn:2023iqa,Higaki:2024jdk}.}
We can combine our scenario with such mass matrices.
Such models would lead to the solution for the strong QCD problem.
In general, the gauge kinetic function has threshold corrections depending 
on other moduli.
How to control effects on $\theta_0$ by other moduli is the next serious issue.

\section{Conclusions}
\label{conclusion}
We have studied moduli stabilization with the motivation of inducing spontaneous CP violation. 
In particular, explicit analyses are performed based on Type IIB string theory on $T^6/\mathbb{Z}_2$ orientifold. 
We take the following two steps to achieve CP violation.
Firstly, background three-form fluxes generate moduli-dependent superpotential, i.e. the Gukov-Vafa-Witten superpotential. 
Along the line of Ref.~\cite{Kobayashi:2020uaj}, we have considered a CP invariant flux-induced potential by restricting the flux quanta. The minimum of the scalar potential does not lead to spontaneous CP violation, instead one finds flat directions where CP-conserving and violating vacua are degenerate. Secondly, we have introduced additional contributions due to matter fields
in the hope of resolving the degeneracy. We have written 
couplings between the moduli and matter fields using modular forms in such a way as to preserve both CP and modular symmetries. 

As a result, the flat directions are lifted and spontaneous CP violation can be realized in some cases. We have presented illustrating examples using $A_4$ modular forms.
We have found that the occurrence of the spontaneous CP violation is dependent on the values of the flux quanta quantized to integers. For instance, if a condition such as eq.~(\ref{eq: CPV_condition}) is satisfied, CP-breaking vacua are obtained quite easily. We have investigated the underlying reasons by focusing on the discrete symmetry of the scalar potential. In general, if a scalar potential is $SL(2,\mathbb{Z}) \rtimes \mathbb{Z}_2^{(\rm CP)}$ invariant, its local minima tend to appear on the CP-conserving points, i.e. the imaginary axis or boundaries of the fundamental domain in the moduli space. Hence, realizing CP violation is not straightforward in such a situation. On the other hand, the modular symmetry of the scalar potential can be modified or broken depending on the flux quanta. 
When the flux quanta satisfy the condition eq.~(\ref{eq: CPV_condition}), the remaining discrete symmetry permits or favors CP violation. Interestingly, the condition (\ref{eq: CPV_condition}) is satisfied under the tadpole cancellation condition only if odd-valued flux quanta are present. This indicates that introducing exotic $O3'$ planes greatly increases the likelihood of spontaneous CP violation. If we relax the tadpole cancellation condition, we can satisfy eq.~(\ref{eq: CPV_condition}) with only even-valued flux quanta.

In both model I and model II of this paper, we have obtained spontaneous CP violation by the VEV of $\tau_2$, one of the three complex structure moduli $\tau_i,\ (i=1,2,3)$ of $T^6/\mathbb{Z}_2$. In model II, we have realized CP violation in the weak coupling regime, $S \sim 2i$ where $S$ denotes the axio-dilaton. The potential minima correspond to the SUSY Minkowski vacua before the K\"{a}hler modulus is stabilized.
For completeness, we have also discussed the K\"{a}hler modulus stabilization through a KKLT-like scenario. 
We have confirmed that it does not ruin our results, namely the realization of spontaneous CP-violation. An intriguing outcome is that the VEV of axio-dilaton can slightly shift and may lead to the solution of the strong QCD problem. In model III which is discussed in Appendix \ref{appendix: Model-III}, we have shown that spontaneous CP violation can be realized by the VEVs of both $\tau_1$ and $\tau_2$ as the de Sitter vacua. We have also found that the modular symmetry of the scalar potential breaks down to the congruence subgroups of $SL(2,\mathbb{Z})$.

Our studies have a direct impact on the modular flavor models whose source of CP violation is the VEVs of the complex structure moduli. Furthermore, the general properties of the CP and modular invariant scalar potential we have discussed are model-independent and, hence largely applicable to the studies on moduli stabilization.

\vspace{1.5 cm}
\noindent
{\large\bf Acknowledgement}\\

We would like to acknowledge the Mainz Institute for Theoretical Physics (MITP) of the Cluster of Excellence PRISMA+ (Project ID 390831469), for enabling us to complete of this work. K.N. would like to thank K.Goto for useful comments. This work was supported in part JSPS KAKENHI Grant Numbers No. JP22K03601 (T.H.),
JP23H04512 (H.O), JP23K03375 (T.K.) and JP24KJ0249 (K.N.).

\appendix
\section*{Appendix}

\section{Condition eq.~(\ref{eq: CPV_condition}) and odd-valued flux quanta}
\label{appendix: CP_condition}

Here, we prove that at least one odd-valued flux quantum needs to be introduced to satisfy the condition eq.~(\ref{eq: CPV_condition}) under the tadpole cancellation condition. We verify this by showing that there exist no integers $m,n$ defined by $|a^0| = 2n,\ |b_3| = 2m$ which simultaneously satisfy the following three requirements:
\begin{enumerate}
    \item $m \nmid n$ and $n \nmid m$.\quad  (cf. (\ref{eq: CPV_condition}) )
\item  $f m n \leq 4$.\quad   (Tadpole cancellation condition, cf. (\ref{eq: tadpole2}))
\item  $nf \in \mathbb{Z},\ mf \in \mathbb{Z}$. \quad (All flux quanta are even integers, cf. (\ref{eq: Restricted_Flux_Quanta}))
\end{enumerate}
We give a proof by contradiction. Let us suppose we have $m, n$ satisfying all the above three requirements.
From the requirement 1, we find
\begin{equation}
    m n \geq 6.
\end{equation}
By combining with the requirement 2, we obtain
\begin{equation}
    f \leq \frac{2}{3}.
\end{equation}
From the requirements 1 and 3, there exist $p, q \in \mathbb{Z}^+$ such that $nf=p,\ mf = q$ where $p \nmid q$ and $q \nmid p$. Then, the following inequality is satisfied
\begin{equation}
    pq = nmf^2 \leq 4f \leq \frac{8}{3} = 2.66\cdots.
\end{equation}
Thus, $pq$ can be either $1$ or $2$. We only have three possibilities for the values of $p$ and $q$,
\begin{equation}
(p,q)=(1,1), (1,2), (2,1).    
\end{equation}
This is a contradiction because $p \mid q$ or $q \mid p$.

\section{Model III}
\label{appendix: Model-III}
Here, we study the following matter K\"{a}hler potential:
\begin{equation}
     K_{\rm matter} = \sum_{I=1,2} Z_I |X_I|^2,
\end{equation}
and the matter superpotential:
\begin{equation}
    W_{\rm matter} = \Lambda^2 (Y_1(\tau_1)  X_1 + \alpha Y_2(\tau_2)X_2), \quad \alpha \in \mathbb{R},
\end{equation}
where
\begin{align}
\begin{aligned}
    Z_1 &= (-i\tau_1 + i \bar{\tau}_1)^{k_1} (-i\tau_2 +i\bar{\tau}_2)^{-1}(-i\tau_3 +i\bar{\tau}_3)^{-1}(-iS +i\bar{S})^{-1}, \\
    Z_2 &= (-i\tau_1 +i\bar{\tau}_1)^{-1}(-i\tau_2 +i\bar{\tau}_2)^{k_2}(-i\tau_3 +i\bar{\tau}_3)^{-1}(-iS +i\bar{S})^{-1}.
\end{aligned}
\end{align} 
The matter field $X_1$ carries modular weights $k_1 \in \mathbb{Z}$ under $\bar{\Gamma}_{\tau_1}$ and $-1$ under $\bar{\Gamma}_{\tau_2}, \bar{\Gamma}_{\tau_3}, \bar{\Gamma}_{S}$. 
The another matter field $X_2$ carries modular weights $k_2 \in \mathbb{Z}$ under ${\bar{\Gamma}}_{\tau_2}$ and $-1$ under $\bar{\Gamma}_{\tau_1}, \bar{\Gamma}_{\tau_3}, \bar{\Gamma}_{S}$.
Then, the trivial singlet modular forms
$Y_1(\tau_1)$ and $Y_2(\tau_2)$ need to carry modular weights $k_{Y1} = -(1+k_1)$ under $\bar{\Gamma}_{\tau_1}$ and $k_{Y2} = -(1+k_2)$ under $\bar{\Gamma}_{\tau_2}$, respectively to preserve the discrete symmetry eq.~(\ref{eq: Full_Modular}).

\subsection{Potential minimum}
Consider the SUSY conditions:
\begin{equation}
    \partial_{\tau_i} W = 0,\quad \partial_S W = 0,\quad  \partial_{X_1} W = 0,\quad \partial_{X_2} W=0,\quad W=0.
\end{equation}
In general, we do not find SUSY-conserving vacua. From $\partial_{\tau_3}W=0$, one obtains 
\begin{equation}
\label{eq: model-III_flat}
   \tau_1 \tau_2 = \frac{b_3}{a^0}.
\end{equation}
On the other hand, $\partial_{X_1}W=0$ and $\partial_{X_2}W=0$ lead to 
\begin{equation}
\label{eq: model-III_tau12}
    Y_1(\tau_1) =0,\quad Y_2(\tau_2) =0.
\end{equation}
We notice that eq.~(\ref{eq: model-III_flat}) and eq.~(\ref{eq: model-III_tau12}) are not compatible in general. Hence, SUSY is spontaneously broken in such cases.

\subsubsection{Scalar potential}
Here, we compute the SUGRA scalar potential of the no-scale type. Assuming $|X_1|, |X_2| \ll 1$, we expand the scalar potential as
\begin{equation}
    V = \sum_{p_1=0}^{\infty} \sum_{q_1=0}^{\infty} \sum_{p_2=0}^{\infty}  \sum_{q_2=0}^{\infty} V_{(p_1,p_2 | q_1, q_2)} X_1^{p_1} \bar{X}_1^{q_1}  X_2^{p_2} \bar{X}_2^{q_2}.
\end{equation}
The leading contribution is given by
\begin{align}
\begin{aligned}
  V_{(0,0|0,0)}&= 
  \frac{\Lambda^4}{\mathcal{V}^2} \left( [2{\rm Im}(\tau_1)]^{k_{Y1}} |Y_1(\tau_1)|^2 + \alpha^2  [2{\rm Im}(\tau_2)]^{k_{Y2}} |Y_2(\tau_2)|^2  \right) \\
  &\quad +  \frac{{2 |\tau_3-fS|^2\  |a^0 \bar{\tau}_1 \tau_2 - b_3|^2 + 2 |\tau_3 - f \bar{S}|^2\  |a^0 \tau_1 \tau_2  - b_3|^2}}{{\mathcal{V}^2 [2{\rm Im}(\tau_1)][2{\rm Im}(\tau_2)][2{\rm Im}(\tau_3)][2{\rm Im}(S)]}}.
  \end{aligned}
\end{align}
If we focus on the regime $\Lambda \ll 1 (=M_{\rm Pl})$, contributions from the three-form fluxes dominate the potential. Hence, there will only be negligible deviations of potential minimum from the flat directions:
\begin{equation}
\label{eq: Constraints_Model_III}
    \tau_1 \tau_2 = \frac{b_3}{a^0},\quad \tau_3 = f S.
\end{equation}
Hence, we are essentially left with the analysis of the following scalar potential:
\begin{equation}
    V_{\rm matter}  = \frac{\Lambda^4}{\mathcal{V}^2} 
    \left( 
    [2{\rm Im}(\tau_1)]^{k_{Y1}} |Y_1(\tau_1)|^2 + \alpha^2 [2{\rm Im}(\tau_2)]^{k_{Y2}} |Y_2(\tau_2)|^2
    \right)
    ,
\end{equation}
under the constraints eq.~(\ref{eq: Constraints_Model_III}). We can regard $V_{\rm matter}$ as a function of only  $\tau_1$ and $\bar{\tau}_1$ by eliminating $\tau_2$ and $\bar{\tau}_2$ as in the Sec.~\ref{subsubsection: Discrete_SYM_Model_I},
\begin{equation}
\label{eq: Matter_Potential_Model_III}
    V_{\rm matter} =  \frac{\Lambda^4}{\mathcal{V}^2}
    \left(
    [2{\rm Im}(\tau_1)]^{k_{Y1}} |Y_1(\tau_1)|^2 + \alpha^2 \left[ \frac{|a^0|}{|b_3|} {2{\rm Im}(\tau_1)} \right]^{k_{Y2}} \left|Y_2 \left(\frac{|a^0|}{|b_3|} \tau_1 \right) \right|^2
    \right).
\end{equation}
We find that $SL(2,\mathbb{Z})_{\tau_1}$ invariance of $V_{\rm matter}$ is broken in general. The discrete symmetry of $V_{\rm matter}$ is given by 
\begin{equation}
\label{eq: Intersection_Discrete_Sym}
 \Gamma^0 ( \tilde{b}_3 ) \cap \Gamma_0 ( \tilde{a}^0 )  = 
 \left\{ 
    \begin{pmatrix}
        a &  b \\
        c & d 
    \end{pmatrix} \in SL(2,\mathbb{Z}) \Bigg| \begin{pmatrix}
        a &  b  \\
        c & d 
    \end{pmatrix} \equiv
   \begin{pmatrix}
            * & \ 0 \ ({\rm mod}~{\tilde{b}_3}) \\
        0 \ ({\rm mod}~{\tilde{a}^0}) & * 
    \end{pmatrix} \right\},
\end{equation} 
where $*$ denotes unspecified integers. We have defined $\tilde{a}^0, \tilde{b}_3 \in \mathbb{Z}^+$ as 
\begin{equation}
    \tilde{a}^0 = \frac{|a^0|}{{\rm gcd}(|a^0|,|b_3|)},\quad   
    \tilde{b}_3 = \frac{|b_3|}{{\rm gcd}(|a^0|,|b_3|)}.
\end{equation}
This can be understood as follows. Recalling the discussions in sec.~\ref{subsubsection: Discrete_SYM_Model_I}, the discrete symmetry of the second term on the right-hand side of eq.~(\ref{eq: Matter_Potential_Model_III}) corresponds to the scale transformed modular group shown by eq.~(\ref{eq: scale_SL2Z}). On the other hand, we find that the symmetry of the first term in eq.~(\ref{eq: Matter_Potential_Model_III}) is $SL(2,\mathbb{Z})$.
Thus, the remaining symmetry (\ref{eq: Intersection_Discrete_Sym})
is given by the intersection between (\ref{eq: scale_SL2Z}) and $SL(2,\mathbb{Z})$.
We briefly explain how the group (\ref{eq: Intersection_Discrete_Sym}) is understood.
An arbitrary element $\gamma'$ in 
the group of (\ref{eq: scale_SL2Z}) is written as
\begin{equation}
\gamma' = 
\begin{pmatrix}
    a' & b' (\tilde{b}_3/\tilde{a}^0) \\
    c'  (\tilde{a}^0/\tilde{b}_3) & d'
\end{pmatrix},
\end{equation}
where $a', b', c'$ and $d'$ are integers, and satisfy $a'd'-b'c'=1$. If one requires $\gamma' \in SL(2, \mathbb{Z})$, there exist integers of $B$ and $C$ such that $b'=\tilde{a}^0 B,\ c' = \tilde{b}_3 C$. Then, we find 
\begin{equation}
\gamma' = 
\begin{pmatrix}
a' & \tilde{b}_3 B \\
\tilde{a}^0 C & d' 
\end{pmatrix} \in \Gamma^0(\tilde{b}_3) \cap \Gamma_0(\tilde{a}^0).
\end{equation}
It is easy to see that an arbitrary element of (\ref{eq: Intersection_Discrete_Sym}) belongs to the intersection between (\ref{eq: scale_SL2Z}) and $SL(2,\mathbb{Z})$. 
Notice that if $\tilde{b}_0=1$ or $\tilde{a}^0=1$, eq.~(\ref{eq: Intersection_Discrete_Sym}) becomes $\Gamma_0(\tilde{a}^0)$ or $\Gamma^0(\tilde{b}_3)$, respectively. If $|a^0|=|b_3|$, it is nothing but $SL(2,\mathbb{Z})$.

\subsection{Examples with $A_4$ modular forms} 
As an illustrating example, we choose
\begin{equation}
    Y_1(\tau_1) = i~Y_{\bf{1}}^{(4)}(\tau_1),\quad  Y_2(\tau_2) = i~Y_{\bf{1}}^{(6)}(\tau_2),
 \end{equation}
where $Y_{\bf{1}}^{(4)}$ and $Y_{\bf{1}}^{(6)}$ denote the $A_4$ trivial singlet modular forms of weight $4$ and $6$, respectively. The overall factors $i$ 
are introduced to satisfy eq.~(\ref{eq: W_matter_CP}).
We have found that when the flux quanta satisfy $b_3 \nmid a^0$ and $a^0 \nmid b_3$, we can realize spontaneous CP violation by the VEVs of both $\tau_1$ and $\tau_2$ quite easily.

\begin{figure}[H]
\centering
 \includegraphics[width=88mm]{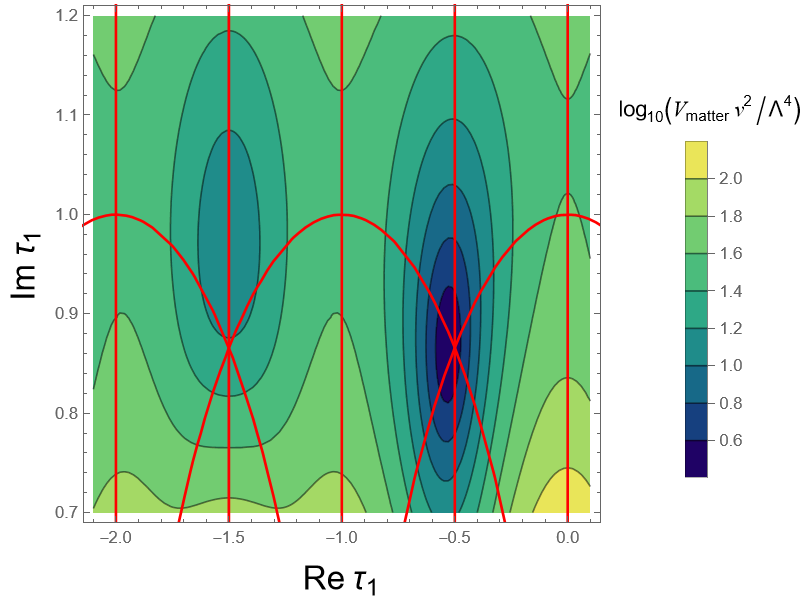}
\caption{Plot of $\log_{10}{(V_{\rm matter}}\mathcal{V}^2/\Lambda^4)$ on the $\tau_1$-plane when $\alpha =0.1$. The matter scalar potential $V_{\rm matter}$ is given by eq.~(\ref{eq: CPV_Example_1}). The red curves denote the CP-conserving region in the moduli space. We have a spontaneous CP violation as the potential minimum is not on the curves.}
\label{fig: CPV_Example_1}
\end{figure}

\paragraph{Example (CP violation):}
Let us choose the following flux quanta:
\begin{equation}
    a^0=2,\quad b_3 = -3,\quad f=1,
\end{equation}
satisfying the tadpole cancellation condition, $n_{\rm flux} = 12 \leq 32$.
We obtain the following scalar potential of the matter contributions:
\begin{equation}
\label{eq: CPV_Example_1}
    V_{\rm matter} = \frac{\Lambda^4}{\mathcal{V}^2} \left( [2{\rm Im}(\tau_1)]^{4}|Y_{\bf{1}}^{(4)}(\tau_1)|^2 + \alpha^2  \left[\frac{4{\rm Im}(\tau_1)}{3}  \right]^{6} \left|Y_{\bf{1}}^{(6)} \left(\frac{2}{3} \tau_1 \right) \right|^2  \right).
\end{equation}
From eq.~(\ref{eq: Intersection_Discrete_Sym}), $V_{\rm matter}$ is symmetric under $\Gamma^0(3) \cap \Gamma_0(2)$ acting on $\tau_1$. The first term of eq.~(\ref{eq: CPV_Example_1}) has a zero-point at $\tau_1 = \omega$ and the second term at $\tau_1 = 3i/2$.
We have shown the contour plot of $\log_{10}{(V_{\rm matter}\mathcal{V}^2/\Lambda^4)}$ on the $\tau_1$-plane when $\alpha = 0.1$ in Fig.~\ref{fig: CPV_Example_1}. The minimum of the potential is located at 
\begin{equation}
\label{eq: CPV_tau1_ModelIII}
   \langle \tau_1  \rangle \simeq  -0.529 + 0.867 i,
\end{equation}
where CP is violated. This corresponds to the de Sitter vacuum because $(V_{\rm matter}\mathcal{V}^2)/\Lambda^4 \simeq 2.82$.
By $T^{-1}S$-transformation, $ \langle \tau_1  \rangle$ is mapped to 
\begin{equation}
    T^{-1} S \langle \tau_1 \rangle \simeq -0.487 + 0.841i,
\end{equation}
which is inside the fundamental domain $\mathcal{D}$.
This is close to the $\mathbb{Z}_3$ fixed point $\tau_1 = \omega$. 
The deviation is given by 
\begin{equation}
\label{eq: ModelIII-tau1_VEV}
    \Delta \tau_1 =  T^{-1} S \langle \tau_1 \rangle - \omega \simeq 0.0286 e^{-1.96 i}.
\end{equation}
We also note that $\tau_2$ is simultaneously stabilized at 
\begin{equation}
    \langle \tau_2 \rangle = - \frac{3}{2\langle \tau_1 \rangle} \simeq 0.770 + 1.26 i,
\end{equation}
where CP is violated. By $T^{-1}$-transformation $\langle \tau_2 \rangle$ is mapped to 
\begin{equation}
    T^{-1} \langle \tau_2 \rangle \simeq -0.230 + 1.26i,
\end{equation}
which is inside the fundamental domain $\mathcal{D}$.

The obtained VEVs are phenomenologically attractive. Firstly, moduli values at the nearby fixed points lead to hierarchical Yukawa couplings which can reproduce the flavor structures of quarks and leptons without fine-tunings \cite{Feruglio:2021dte,Novichkov:2021evw,Petcov:2022fjf,Kikuchi:2023cap,Abe:2023ilq,Kikuchi:2023jap,Abe:2023qmr,Petcov:2023vws,Abe:2023dvr,deMedeirosVarzielas:2023crv,Kikuchi:2023dow,Kikuchi:2023fpl}. 
Typically, deviations of order $|\langle \tau \rangle - \tau_{*}| \sim \mathcal{O}(0.01-0.1)$ are preferred where $\tau_{*}$ denotes the finite fixed points.
We have realized such a small deviation by the VEV of $\tau_1$ as indicated by eq.~(\ref{eq: ModelIII-tau1_VEV}). Secondly, 
we can realize both sizable CP violation and mass hierarchies quite well if more than one moduli VEVs break CP.
In a modular flavor model with a single modulus, it is difficult to reproduce sufficient CP violation unless explicit CP-breakings are assumed at the tree level \cite{Petcov:2022fjf,Kikuchi:2023cap,Abe:2023ilq,Kikuchi:2023jap,Abe:2023qmr,Petcov:2023vws,Abe:2023dvr,deMedeirosVarzielas:2023crv,Kikuchi:2023dow,Kikuchi:2023fpl}. We have succeeded in generating CP violation by the VEVs of two moduli $\tau_1$ and $\tau_2$.

\begin{figure}[H]
\centering
 \includegraphics[width=88mm]{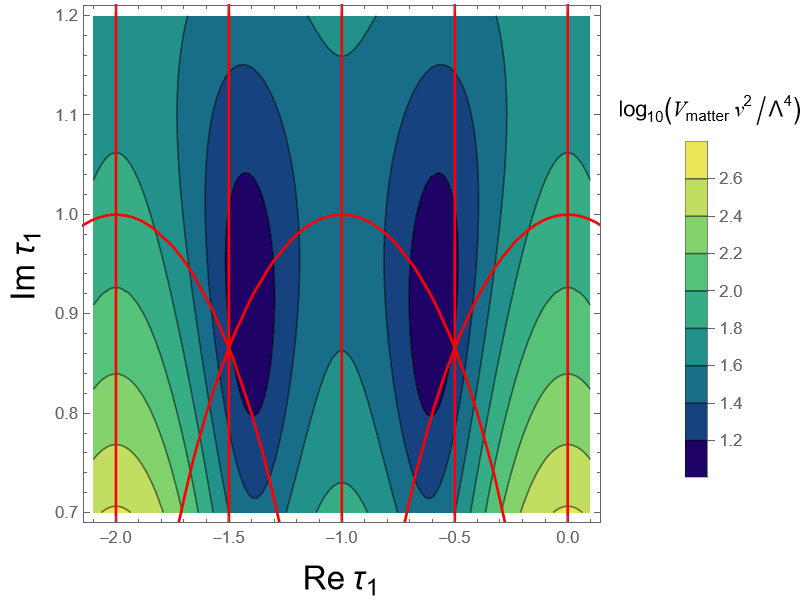}
\caption{Plot of $\log_{10}{(V_{\rm matter}\mathcal{V}^2}/\Lambda^4)$ on the $\tau_1$-plane when $\alpha =0.1$ and $b_3/a^0=-2$.
 Red curves denote the CP-conserving region in the moduli space. We have checked that CP is conserved because the potential minima precisely lie on the curves.}
\label{fig: NoCPV_Example}
\end{figure}

\paragraph{Example (No CP violation):}
Let us present one more example. We choose the following flux quanta:
\begin{equation}
    a^0=2,\quad b_3 = -4,\quad f=1,
\end{equation}
satisfying the tadpole cancellation condition $n_{\rm flux} = 16 \leq 32$.
We obtain the following scalar potential of the matter contributions:
\begin{equation}
\label{eq: No_CPV_Example_1}
    V_{\rm matter} = \frac{\Lambda^4}{\mathcal{V}^2} \left( [2{\rm Im}(\tau_1)]^{4}|Y_{\bf{1}}^{(4)}(\tau_1)|^2 + \alpha^2  \left[ {\rm Im}(\tau_1)  \right]^{6} \left|Y_{\bf{1}}^{(6)} \left(\frac{1}{2} \tau_1 \right) \right|^2  \right).
\end{equation}
From eq.~(\ref{eq: Intersection_Discrete_Sym}), $V_{\rm matter}$ is symmetric under $\Gamma^0(2)$ acting on $\tau_1$. The first term of eq.~(\ref{eq: No_CPV_Example_1}) has a zero-point at $\tau_1 = \omega$ and the second term at $\tau_1 = 2i$.
We have shown the 
contour plot of $\log_{10}{(V_{\rm matter}\mathcal{V}^2/\Lambda^4)}$ on the $\tau_1$-plane when $\alpha = 0.1$ in Fig.~\ref{fig: NoCPV_Example}. We do not obtain CP violation because the minimum in Fig.~\ref{fig: NoCPV_Example} lies on a CP invariant arc, i.e. $|\tau_1+1|=1$. We have numerically determined the potential minimum:
\begin{equation}
\label{eq: tau1_VEV_No_CPV}
   \langle \tau_1  \rangle \simeq  -0.591 + 0.912i.
\end{equation}
The minimum is precisely on the CP-conserving arc owing to the residual symmetry $\Gamma^0(2)$ as discussed in Appendix 
\ref{appendix: Gamma_0(a)_Gamma^(b)}. 
We note that $\tau_2$ is simultaneously stabilized at 
\begin{equation}
    \langle \tau_2 \rangle = - \frac{2}{\langle \tau_1 \rangle} \simeq 1.00 +
1.54i,
\end{equation}
where CP is conserved. If $\langle \tau_1 \rangle$ is on the CP invariant arc $\tau_1=-1+e^{i \theta}$, we obtain ${\rm Re}\langle \tau_2\rangle = - {\rm Re}( 2/\langle \tau_1 \rangle) = 1$ regardless of the value of $\theta \in \mathbb{R}$. This implies that CP violation is hardly realized even when $\alpha$ is slightly shifted from $0.1$.

In our model, we have assumed $\Lambda \ll M_{\rm Pl}$ and assumed that deviations from the flat directions (\ref{eq: Constraints_Model_III}) are negligible. On the other hand, if we consider the regime where $\Lambda$ becomes comparable to $M_{\rm Pl}$, such deviations may generate CP violation and
need to be evaluated carefully. We leave the comprehensive studies for future work.

\subsection{CP and $\Gamma_0(a) \cap \Gamma^0(b)$ invariance}
\label{appendix: Gamma_0(a)_Gamma^(b)}
Here, we study the implications of the symmetry of the scalar potential, cf. eq.~(\ref{eq: Intersection_Discrete_Sym}).
Let us consider the following two elements in $\Gamma_0(a) \cap \Gamma^0(b)$,
\begin{equation}
T^b = \begin{pmatrix}
        1 & b \\ 0 & 1
    \end{pmatrix},\quad 
    ST^{-a}S^{-1} =
    \begin{pmatrix}
       1 & 0 \\ a &  1 
    \end{pmatrix},
\end{equation}
where $a, b$ are positive integers. $T^b$ symmetry will just be the same as before. Hence, let us focus on the $ST^{-a}S^{-1}$ invariance. 
It is convenient to parametrize the modulus $\tau$ which 
as 
\begin{equation}
    \tau=-\frac{1}{a} + \Delta \tau ,\quad \Delta \tau = r e^{i \theta}.
\end{equation}
As in Sec.~\ref{sec:Gamma0(N)}, we find 
\begin{equation}
    V(r, \theta) = V(1/(a^2r), \theta),
\end{equation}
which leads to 
\begin{equation}
    \frac{\partial V}{\partial r}(r=1/a, \theta) = 0.
\end{equation}
This shows that for $a = 1$,
the derivative of $V$ does vanish on the CP-conserving unit arcs including $|\tau + 1|=1$ as we commented just after eq.~(\ref{eq: tau1_VEV_No_CPV}). On the other hand, for $a \neq 1$, it will not vanish on the unit arcs.
We consider this allows
the slight deviation of the modulus VEV from the CP invariant arc in the first example, i.e. (\ref{eq: CPV_tau1_ModelIII}).

\section{Modular forms of $A_4$}
\label{appendix: modular_A4}
Here we review the modular forms of $A_4$. They can be constructed in terms of the Dedekind eta function and its first derivative:
\begin{align}
  &\eta(\tau) = q^{1/24} \prod_{n=1}^{\infty} (1-q^n), \quad q = e^{2\pi i\tau}, \\
  &\eta'(\tau) \equiv \frac{d}{d\tau} \eta(\tau),
\end{align}
where $\tau$ is a complex number with positive imaginary part, i.e. ${\rm Im}(\tau) > 0$. In Ref.~\cite{Feruglio:2017spp}, $A_4$ triplet $\bf{3}$ modular forms of weight $2$ have been constructed as
\begin{align}
  Y^{(2)}_{\bm{3}}(\tau) = 
  \begin{pmatrix}
    Y_1 \\ Y_2 \\ Y_3 \\
  \end{pmatrix},
\end{align}
where
\begin{align}
  &Y_1(\tau) = \frac{i}{2\pi} \left(\frac{\eta'(\tau/3)}{\eta(\tau/3)} + \frac{\eta'((\tau+1)/3)}{\eta((\tau+1)/3)} + \frac{\eta'((\tau+2)/3)}{\eta((\tau+2)/3)} - \frac{27\eta'(3\tau)}{\eta(3\tau)}\right), \\
  &Y_2(\tau) = \frac{-i}{\pi} \left(\frac{\eta'(\tau/3)}{\eta(\tau/3)} + \omega^2\frac{\eta'((\tau+1)/3)}{\eta((\tau+1)/3)} + \omega \frac{\eta'((\tau+2)/3)}{\eta((\tau+2)/3)}\right), \\
  &Y_3(\tau) = \frac{-i}{\pi} \left(\frac{\eta'(\tau/3)}{\eta(\tau/3)} + \omega\frac{\eta'((\tau+1)/3)}{\eta((\tau+1)/3)} + \omega^2\frac{\eta'((\tau+2)/3)}{\eta((\tau+2)/3)}\right).
\end{align}
By taking tensor products of $Y_i(\tau)$, we can construct $A_4$ modular forms of higher weights. For example, trivial singlet $A_4$ modular forms of weight $4$ and $6$ are given by  
\begin{align}
\label{eq: A4_trivial_singlet_Y}
  Y_{\bm{1}}^{(4)}(\tau) = Y^2_1+2Y_2Y_3,
\end{align}
and
\begin{align}
  Y^{(6)}_{\bm{1}}(\tau) = Y^3_1+Y^3_2+Y^3_3-3Y_1Y_2Y_3,
\end{align}
respectively. Their $q$-expansions are given by \cite{Abe:2024tox}
\begin{align}
     Y_{\bm{1}}^{(4)}(\tau) &= 1+240 q+2160 q^2+6720q^3+\mathcal{O}(q^4), \\
     Y^{(6)}_{\bm{1}}(\tau) &= 1- 504 q -16632 q^2 - 122976 q^3 +\mathcal{O}(q^4).
\end{align}


 \end{document}